\documentclass[12pt]{article}
\usepackage{graphicx}
\usepackage{bm}
\def\be{\begin{equation}}
\def\ee{\end{equation}}
\def\bea{\begin{eqnarray}}
\def\eea{\end{eqnarray}}
\usepackage{latexsym}
\usepackage{dcolumn}
\usepackage{epsfig,amssymb,euscript,amsfonts}
\usepackage{amsmath}
\usepackage{comment}
\usepackage{array,calc,epsfig}

\usepackage[american]{babel}
\usepackage{color}
\usepackage{hyperref}
\usepackage{enumerate}
\usepackage{float}
\usepackage{enumitem}

\usepackage{xcolor}
\usepackage{subcaption}
\usepackage{chngcntr}
\usepackage[noadjust]{cite}

\def\be{\begin{equation}}
\def\ee{\end{equation}}
\def\ba{\begin{eqnarray}}
\def\ea{\end{eqnarray}}
\def\nb{\nonumber}
\def\p{\partial}

\def\d{\delta}

\def\o{\omega}
\def\O{\Omega}
\def\r{\rho}
\def\s{\sigma}

\def\q{\quad}

\def\mc{\mathcal}

\newcommand{\pr}[1]{\left(#1\right)}
\newcommand{\pq}[1]{\left[#1\right]}

\counterwithin*{equation}{section}

\usepackage[normalem]{ulem}

\usepackage{color}

\begin{document}
\baselineskip=15.5pt
\pagestyle{plain}
\setcounter{page}{1}
\newfont{\namefont}{cmr10}
\newfont{\addfont}{cmti7 scaled 1440}
\newfont{\boldmathfont}{cmbx10}
\newfont{\headfontb}{cmbx10 scaled 1728}
\renewcommand{\theequation}{{\rm\thesection.\arabic{equation}}}
\renewcommand{\thefootnote}{\arabic{footnote}}

\vspace{1cm}
\begin{titlepage}
\vskip 2cm
\begin{center}
{\Large{\bf Bubble wall velocity and nucleation rates\\ in inverse holographic phase
transitions}}
\end{center}
\vskip 10pt
\begin{center}
Francesco Bigazzi$^{a}$, Aldo L. Cotrone$^{a,b}$, Natalia Pinzani-Fokeeva$^{a}$, Tommaso Trabocchi$^{b}$
\end{center}
\vskip 10pt
\begin{center}
\vspace{0.2cm}
\textit {$^a$ INFN, Sezione di Firenze; Via G. Sansone 1; I-50019 Sesto Fiorentino (Firenze), Italy.
}\\
\textit{$^b$ Dipartimento di Fisica e Astronomia, Universit\'a di Firenze; Via G. Sansone 1;\\ I-50019 Sesto Fiorentino (Firenze), Italy.
}\\
\vskip 20pt
{\small{bigazzi@fi.infn.it, cotrone@fi.infn.it, n.pinzanifokeeva@gmail.com, tommaso.trabocchi@edu.unifi.it}}

\end{center}

\vspace{25pt}

\begin{center}
 \textbf{Abstract}
\end{center}

\noindent 
We study the dynamics of first-order inverse phase transitions (driven by superheating) at strong coupling, focusing on the top-down Witten-Sakai-Sugimoto model for holographic QCD. Two cases are considered: the deconfinement transition in the unflavored version of the model and a chiral symmetry-restoring transition occurring in the deconfined phase of the full theory. In both cases, we imagine driving the system into a metastable phase at high temperature, inducing the nucleation of bubbles of the stable phase. For both classes of transitions, we find the corresponding Euclidean bounce solutions and compute the bubble nucleation rates and the relevant transition parameters. For the deconfinement transition, the large jump in the number of degrees of freedom between the two phases suggests that the bubble wall velocity is parametrically small; we provide a rough estimate of it near the critical temperature. In the case of the chiral transition, instead, we compute the bubble wall velocity and the friction force exerted on the bubbles 
employing motivated ansatze and approximations for the steady-state configurations.

\end{titlepage}
\newpage
\tableofcontents
\section{Introduction}

Cosmological {\it direct} first-order phase transitions, occurring in dark matter or beyond the Standard Model sectors and driven by the expansion and cooling of the Universe, have attracted considerable attention in recent years. A primary motivation is their potential to generate  stochastic gravitational waves that may be detectable by current and future experiments, providing a window into new physics. See e.g.~\cite{Caprini:2015zlo,Caprini:2019egz,Hindmarsh:2020hop,Athron:2023xlk} for reviews. 

First-order phase transitions are, in fact, violent dynamical processes that proceed via the nucleation of bubbles of a new phase within the metastable old phase, when a parameter, such as the temperature, is varied sufficiently far from criticality. When the  nucleation rate becomes large enough, bubbles nucleate, expand, collide, and eventually percolate, completing the transition. In cosmological contexts, gravitational waves may then be generated through bubble collisions, sound waves in the surrounding plasma, and magnetohydrodynamic turbulence. 

The gravitational-wave spectrum depends on the details of the underlying model, which might be weakly or strongly coupled, and, in particular, on several key parameters  such as the nucleation and percolation temperatures $T_n$ and $T_p$, the transition strength $\alpha$, the inverse transition duration $\beta$, and the steady state bubble-wall velocity $v$, which is a challenging out-of-equilibrium parameter to compute even at weak coupling. It is therefore of considerable importance to determine these quantities in view of forthcoming gravitational-wave observations. The holographic correspondence emerges as a unique tool for this purpose when the underlying physics is strongly coupled. See e.g.~\cite{Bigazzi:2020phm,Bigazzi:2020avc,Ares:2020lbt,Bea:2021zsu,Bigazzi:2021ucw,Ares:2021ntv,Bea:2022mfb,Hoyos:2026eoq}.

Recently, several works have drawn attention towards {\it inverse} first-order phase transitions \cite{Barni:2024lkj,Bea:2024bxu} driven by superheating. While these transitions might be relevant in cosmological contexts, such as transient phenomena like reheating phases, they might also play a prominent role in other astrophysical settings, such as the evolution of supernovae and neutron star mergers, which involve high-density QCD physics at strong coupling. For the same reasons as reviewed before, these transitions, if realized, might induce the production of gravitational waves. 

Neutron stars and their mergers, as well as supernovae, may actually provide a complementary probe, alongside heavy-ion collision experiments, of the QCD phase diagram (see e.g.~\cite{Borsanyi:2025ttb} for a short recent review), offering a unique window on its still unknown features at large enough baryon chemical potential. In particular, numerical simulations indicate that when neutron stars merge, they may form lumps of hot, dense matter in which a quark-matter phase may become thermodynamically favored, even if the cores of the original stars were in a hadronic phase; see, e.g., \cite{Most:2018eaw}. If the temperature is large enough or the matter is sufficiently compressed, a transition from  hadronic matter to quark matter may thus occur in these regions. 
It is currently not known whether this is a first order transition at large enough density or not, even though this is expected from many effective models. 
If the transition is of first order, it may generate both a post-merger gravitational-wave signal 
different from that expected without the transition, due to a difference in the equation of state,
and a stochastic gravitational-wave background in the MHz region \cite{Blas:2022xco}. 

The dynamical evolution of direct and inverse phase transitions is not symmetric \cite{Barni:2024lkj,Bea:2024bxu}. Differences in the transport coefficients and densities of the two phases, as well as the direction of heat flow, affect the hydrodynamic response of the surrounding medium.\footnote{For instance, in a more familiar non-relativistic example, a vapor bubble absorbs heat from the surrounding liquid to sustain its growth, while a liquid droplet releases latent heat into the surrounding vapor phase during condensation. Moreover, because the two phases have different densities and viscosities, the hydrodynamic response of the surrounding medium differs significantly between the two cases.} As a consequence, the steady-state bubble-wall velocity reached at late times and the friction force exerted on the bubbles are expected to show differences between the direct and inverse transitions. For the bubble wall velocity, this has been actually observed in the case of a chiral-like phase transition in the deconfined phase of a holographic bottom-up model \cite{Bea:2024bls}.
For other recent studies of inverse phase transitions see \cite{Barni:2025gnm,Barni:2025mud,Barni:2025ced,Bleau:2026ala}.

With the aim of describing the dynamics of inverse phase transitions at strong coupling, in this work we will focus on the Witten-Sakai-Sugimoto (WSS) model of holographic QCD \cite{Witten:1998zw,Sakai:2004cn}. The model provides a holographic dual description of a strongly coupled, large-$N$ $SU(N)$ gauge theory, coupled to $N_f\ll N$ quarks (and to a tower of massive adjoint Kaluza-Klein fields), that shares several qualitative features with QCD at low energy. The model parameters can be adjusted to either mimic QCD physics or to describe more general QCD-like strongly coupled dark matter sectors. 

The WSS model exhibits two classes of first-order phase transitions at finite temperature: the confinement-deconfinement transition and chiral symmetry-breaking and-restoring transitions. Whilst in the confined phase chiral symmetry is always broken, in the deconfined phase it can be either broken or preserved depending on the model parameters \cite{Aharony:2006da}. The phase diagram of the WSS model is enriched at finite (e.g.~baryon) chemical potential or in the presence of external (e.g.~magnetic) fields. As a first step, we will begin by turning off these extra ingredients, leaving their analysis for the future.

The confinement-deconfinement phase transition involves a large jump in the entropy density between the two phases, which, in the 't Hooft limit, is dominated by the Yang-Mills sector of the model. For this reason, when we focus on this transition, we will neglect the flavor sector. The chiral phase transition in the model's deconfined phase, by contrast, does not exhibit such a large jump: the physics is dominated by the flavor sector, with the Yang-Mills plasma as a background. 
In this sense, it provides an interesting relative of the expected transition from baryonic to quarkionic matter in QCD at large density.

The relevant parameters for the direct supercooled transitions occurring as the system cools down from a high-temperature phase to a low-temperature one — namely, from the deconfined to the confined phase and from the chirally restored  to the chirally broken phase — have been computed in \cite{Bigazzi:2020phm} and  \cite{Bigazzi:2021ucw}. 

Here, instead, we consider the inverse superheated process. We will thus heat up the WSS model from the confined to the deconfined phase and from the chirally broken to the chirally restored phase. Each transition will proceed through the nucleation of deconfined or chirally restored bubbles 
within a metastable superheated phase.
We will then compute several parameters characterizing these transitions. In particular, for both the deconfinement and the chiral symmetry restoring transition, we will find the Euclidean ``bounce" solutions  \cite{Coleman:1977py,Linde:1980tt} interpolating between the metastable and the stable phase. We will then compute the bubble nucleation rates and other relevant transition parameters. 

In the case of the deconfinement transition, following a suggestion in \cite{Sanchez-Garitaonandia:2023zqz}, viable for transitions involving a large jump in the number of degrees of freedom, we will observe that the bubble wall velocity might be parametrically very small in the holographic limit.  We will just provide a rough estimate of it for temperatures very close to the critical one.

For the chirally restoring phase transition, instead, we will be able to provide a first principle result on both the steady-state bubble wall velocity (eq. \eqref{equv} being one of the central results of this work) and the friction force exerted on the bubble, at least within certain reasonable assumptions and approximations of the steady-state configuration for a thin-wall bubble in dynamical equilibrium with the surrounding plasma as in \cite{Bigazzi:2021ucw}. We will show that our results, and the related differences with the supercooled case studied in \cite{Bigazzi:2021ucw}, are in qualitative agreement with expectations from \cite{Barni:2024lkj,Bea:2024bxu} and with the results on the bubble-wall velocity in \cite{Bea:2024bls}.
Nevertheless, the top-down WSS model admits much larger velocities than those obtained in the bottom-up model analyzed in \cite{Bea:2024bls}.
Moreover, it allows one to realize that it is the enthalpy density of the \emph{true vacuum} that provides the resistance to the wall motion, in contrast to what happens in the corresponding supercooled transition, where the same role is played by the enthalpy density of the \emph{false vacuum}. 

Our work is organized as follows. First, we review the WSS model in section \ref{S:WSS}. Then, we study  the nucleation of deconfined vacuum bubbles in a metastable phase, where the theory is confining, and we estimate the corresponding steady-state bubble wall velocity in section \ref{S:dec}. Then, we consider the nucleation of bubbles of chirally symmetric vacuum in a plasma where chiral symmetry is broken  in section \ref{eq:DBI}. Finally, we compute the steady-state velocity and the friction force of these bubbles in section \ref{S:vel}. We end with a conclusion section \ref{S:concl}.

\section{The holographic model}
\label{S:WSS}
The model we will focus on, in its low-energy regime, is 
a large-$N$ QCD-like theory with a top-down string embedding. 
The Yang-Mills sector arises from the low energy dynamics of $N$ $D4$-branes wrapped on a circle $x_4\sim x_4+2\pi M_{KK}^{-1}$ around which bosons (resp.~fermions) are taken to have periodic (resp.~anti-periodic) boundary conditions \cite{Witten:1998zw}. At energies $E\ll M_{KK}$ it is a $3+1$ dimensional $SU(N)$ non-supersymmetric gauge theory, coupled to adjoint Kaluza-Klein fields whose mass scale is set by $M_{KK}$. Quarks, i.e.~chiral fundamental fermionic matter fields, are added by means of two stacks of $N_f$ $D8$ and $N_f$ $\bar{D}8$-branes placed at distance $L\le \pi M_{KK}^{-1}$ in the $x_4$ direction \cite{Sakai:2004cn}. 

We will consider massless quarks and focus on the $N_f\ll N$ regime of the model, where quantum effects due to quark loops are subleading and, correspondingly, the backreaction of the $D8-\bar{D}8$-branes on the holographic background can be neglected. The latter is the near-horizon limit of the background sourced by the $D4$-branes. The corresponding classical gravity solution is reliable if $N\gg1$ and $\lambda\gg1$, where $\lambda=4\pi g_s N M_{KK} l_s$ is a 't Hooft-like parameter, which turns out to be  proportional to the ratio between the confining string tension of the dual gauge theory and $M_{KK}^2$. 

The confining phase of the model is dual to a ten-dimensional type IIA soliton-like background sourced by the $D4$-branes. The string frame metric (in Euclidean signature) reads
\bea
\label{solitme}
ds^2 &=& \pr{\frac{u}{R}}^{3/2} \pq{dt^2 + dx_i dx_i +  f(u) d x_4 ^2 } +\pr{\frac{R}{u}}^{3/2} \pq{\frac{du^2}{f_T(u)} +u^2 
d \O _4 ^2} \,,\nonumber \\
f(u)&=&1-\frac{u_0^3}{u^3}\,,\quad u_{0}=\frac{4}{9} R^3 M_{KK}^2 \,,\quad R^3=\pi g_s N l_s ^3\,.
\eea
Here $t\sim t+1/T$ and $x^i$ span the directions of the dual gauge theory at temperature $T$. The background also features a four-sphere $S^4$ and a regular cigar-shaped subspace with disk topology, spanned by the coordinates $ x_4$ and $ u$. Here $u\in [u_0,\infty)$ is the holographic radial coordinate dual to the renormalization group energy scale of the gauge theory. Regularity at the tip of the cigar enforces the specific relation between $u_0$ and $M_{KK}$ shown above. The background includes a running dilaton and a RR four-form $F_4$ whose flux along the $S^4$ is proportional to $N$
\be
e^\phi=g_s\left(\frac{u}{R}\right)^{3/4}\,,\quad F_4= \frac{3 R^3 }{g_s} \o_4\,,
\ee
where $\o_4$ is the volume form of $S^4$.

The $D8-\bar{D}8$ branes form a non-trivial U-shaped profile in the above background. This geometrically realizes the $U(N_f)_R\times U(N_f)_L\rightarrow U(N_f)_D$ breaking of chiral symmetry in the dual gauge theory. The position $u=u_J$ of the tip of the connected configuration sets the chiral symmetry breaking scale $f_{\chi}$ and depends on the asymptotic $D8-\bar{D}8$ distance $L$. In the antipodal case where $L=\pi M_{KK}$, $u_J=u_0$ so that the scale of confinement is the same as that of chiral symmetry breaking.  

The deconfined phase of the model is dual to a planar black-hole background with string frame metric (in Euclidean signature)
\bea
\label{bhme}
ds^2 &=& \pr{\frac{u}{R}}^{3/2} \pq{f_T(u)dt^2 + dx_i dx_i +  d x_4 ^2 } +\pr{\frac{R}{u}}^{3/2} \pq{\frac{du^2}{f_T(u)} +u^2 
d \O _4 ^2} \,,\nonumber \\
f_T(u)&=&1-\frac{u_T^3}{u^3}\,,\quad u_{T}=\frac{16\pi^2}{9} R^3 T^2 \,,\quad R^3=\pi g_s N l_s ^3\,.
\eea
The coordinate $u$ now runs in $[u_T,\infty)$, where $u_T$ is the radial position of the event horizon. The subspace $(x_4,u)$ now has the topology of a cylinder. Regularity of the Euclidean solution enforces the relation between $u_T$ and the temperature $T$ shown above. The dilaton and the $F_4$ field remain the same as in the solitonic case. Notably, the black-hole background can be obtained from the solitonic one by exchanging $x_4\leftrightarrow t$ and $u_0\leftrightarrow u_T$. 

The holographic description entails the presence of a first-order phase transition between the confined and deconfined phases when the temperature exceeds the critical temperature $T_c = M_{KK}/2\pi$. 
Moreover, in the deconfined phase (at $T>M_{KK}/2\pi$), another chiral first-order phase transition may occur. In this case, a disconnected $D8-\bar{D}8$ configuration, with the branes stretched along $u$ and reaching the event horizon, is always allowed, and corresponds to a chirally symmetric phase. However, if $M_{KK}/2\pi<T<0.1538 L^{-1}$ a chirally broken U-shaped configuration is energetically preferred. A first order phase transition with critical temperature $T_c^{\chi}\approx0.1538 L^{-1}$ occurs between the chirally broken and the chirally symmetric phase \cite{Aharony:2006da}.

\section{Deconfinement transition}
\label{S:dec}
In this section, we will focus on the first-order confinement-deconfinement transition of the unflavored version ($N_f=0$) of the holographic model reviewed above. In particular, we will assume to start in the confined phase and then heat up the model above $T_c$. In this case, bubbles of the new preferred (deconfined) phase are expected to nucleate within the metastable confined one. The process we have in mind is thus the superheated version of the supercooled case examined in the first part of \cite{Bigazzi:2020phm}, to which we refer for most of the background results. 

To describe the nucleation of bubbles, a ten-dimensional solution interpolating (in space) between the confined  and the deconfined backgrounds  (\ref{solitme}) and (\ref{bhme}) would be needed. As this is a very challenging task, in \cite{Bigazzi:2020phm} a simpler, effective approach was adopted, where, following the suggestions in \cite{Creminelli:2001th}, the interpolation was mediated by a single effective degree of freedom. Taking advantage of the already mentioned map between the two backgrounds, the idea was to promote the parameters $u_T$ and $u_0$ to fields $u_T(\r)$ and $u_0(\r)$, and in turn to a piece-wise defined scalar $\Phi(\rho)$, where $\r$ is the radial coordinate of the bubble. As we are interested in the superheated case, here we will focus on $O(3)$ symmetric bubbles, for which $\rho^2=x_i x_i$. 

The effective action for the bubbles was found in \cite{Bigazzi:2020phm} to be given by 
\be
\frac{S_3(\Phi)}{T}=\frac{32\pi^4 g}{3^5 \bar T} \int_{0}^{\infty}d\bar\rho \bar\rho^2 \left [ \left(5-\frac{\pi}{2\sqrt{3}} \right) \Phi'^2 +  \Theta(\Phi) V_{R}(\Phi) + \Theta(-\Phi) V_{L}(\Phi) \right]\,,
\label{o3dc}
\ee
where $\Theta(\cdot)$ is the Heaviside step function, 
\bea
V_{R}(\Phi) &=& \frac{16\pi^2}{9}\left(5\Phi^3-\frac{3}{\pi}\Phi^{5/2}\right)\,,\nonumber \\
V_{L}(\Phi) &=& -\frac{16\pi^2}{9}\left(5\Phi^3+\frac{3}{\pi}\bar T (-\Phi)^{5/2}\right)\,,
\eea
and
\be
g \equiv \lambda N^2\,,\quad \bar \rho \equiv M_{KK} \rho\,, \quad \bar T \equiv \frac{2\pi T}{M_{KK}}=\frac{T}{T_c}\,.
\ee
When $\Phi<0$ (resp.~$\Phi>0$), the phase of the system is in a deconfined (resp.~confined) branch. 

\begin{figure}
\center
\includegraphics[scale=0.7]{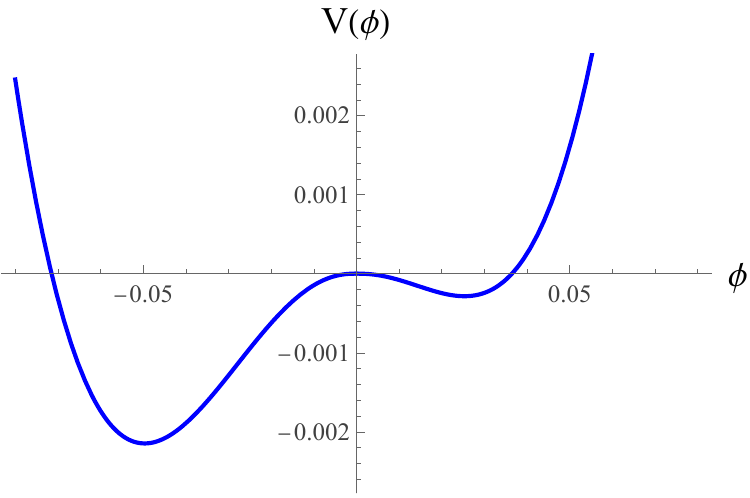}
\caption{The potential for $\bar T=1.4$.}
\label{figpot}
\end{figure}

The potential has two minima $V_D=-\bar T^6/(36 \pi^4)$ for $\Phi_D=-\bar T^2/(4\pi^2)$ and $V_C=-1/(36 \pi^4)$ for $\Phi_C= 1/(4\pi^2)$. These correspond, respectively, to the deconfined and the confined solutions reviewed above. We will focus on the case $\bar T >1$, where the true vacuum is the deconfined one at $\Phi=\Phi_D$. See figure \ref{figpot} for an explicit example. 

The bubble-like bounce solution $\Phi_B$ of the Euler-Lagrange (EL) equation derived from the action (\ref{o3dc}) is obtained as follows \cite{Coleman:1977py}.
Inside the bubble, i.e.~for $\bar\rho \in [0,\bar\rho_w]$ where $\bar\rho_w$ is the location of the bubble wall (that is the radius of the bubble),  we want to find a deconfined phase with $\Phi<0$. To do so, we solve the EL equation with boundary conditions
\be
\Phi _B (0)=\Phi_0 \ , \q \q \q \Phi' _B(0)=0 \ ,
\ee
for some negative value $\Phi_0$; regularity fixes the second condition above.
The solution $\Phi _B$ will vanish at some finite value of the radius, identified with $\bar\rho_w$, with derivative $\Phi' _B(\bar\rho_w)\equiv \Phi'_{B, w}$.

We then solve the equation outside the bubble, i.e.~for $\bar\rho \in [\bar\rho_w, \infty]$, where now we want to find the system in the confined branch where $\Phi>0$. 
Enforcing continuity of $\Phi _B$ and $\Phi' _B$ we now require 
\be
\Phi _B(\bar\rho_w)=0 \ , \q \q \q \Phi' _B(\bar\rho_w)=\Phi'_{B,w} \ .
\ee
Finally, we look for the initial value $\Phi_0$ at the center of the bubble such that the solution for large $\bar\rho$ goes to the false vacuum $\Phi_C$,
\be
\Phi_B(\infty)=\Phi_C\,.
\ee
The whole bounce solution is such that, at the center of the bubble, it goes to a negative constant\footnote{The constant $\Phi _0$ is typically different from the true vacuum $\Phi_D$, because the equation of motion derived from (\ref{o3dc}) contains a friction term.} with vanishing derivative, and at infinity it goes to the false vacuum solution. 
\begin{figure}
\center
\includegraphics[scale=0.7]{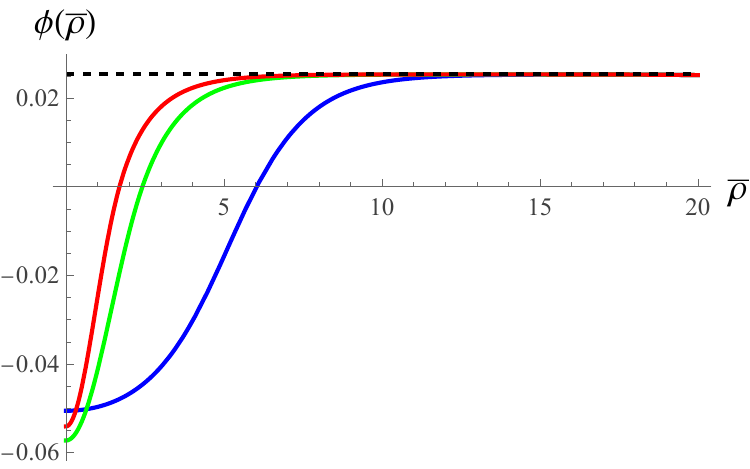}
\caption{Solutions for the bubble profile with $\bar T=1.5$ (Blue line), $\bar T=3$ (Green line), $\Bar T=5$ (Red line). The horizontal dashed line corresponds to the (constant) value of the field in the confining (false) vacuum.}
 \label{figsol}
\end{figure}
Examples of solutions corresponding to different choices of $\bar T$ are given in figure \ref{figsol}.
The bubble wall radius increases as we approach the critical temperature, consistently with expectations from the thin-wall approximation; it then decreases as the temperature increases. This is shown in figure \ref{bwr}. The wall radius is always greater than the radius $\rho_T= (2\pi T)^{-1}$ of the thermal circle, confirming the validity of the $O(3)$ symmetric configuration. For large enough temperature, data on the bubble wall radius are quite well approximated by the function $\bar\rho_w= 3.5 (\bar T-1)^{-1/2}$.
\begin{figure}
\center
\includegraphics[scale=0.7]{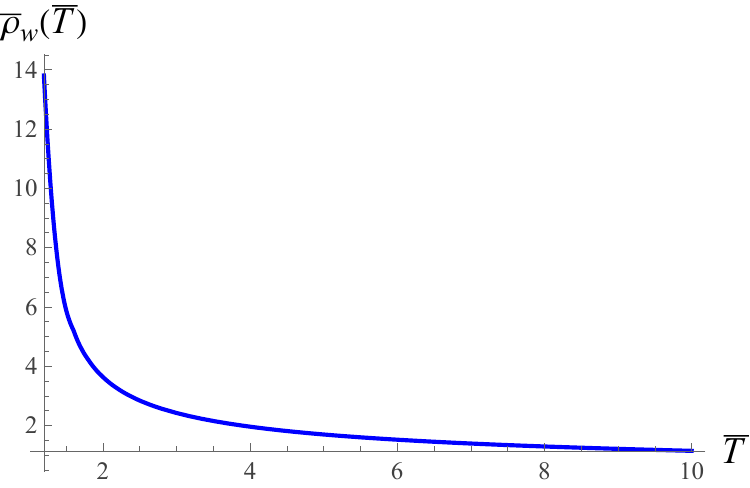}
\caption{Bubble wall radius as a function of $\bar T$.}
 \label{bwr}
\end{figure}
\subsection{Bubble nucleation rate}
\begin{figure}
\center
\includegraphics[scale=0.7]{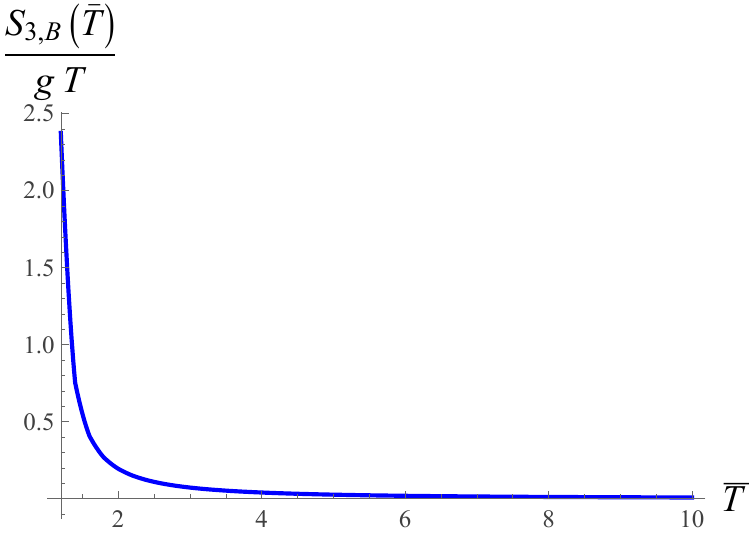}
\caption{Bounce action (divided by the coupling $g$) as a function of $\bar T$.}
 \label{bact}
\end{figure}

The nucleation rate of the bubbles can be estimated by first computing the so-called bounce action \cite{Coleman:1977py,Linde:1980tt}
\be
\label{eq:bounce.action}
\frac{S_{3,B}}{T} = \frac{S_3(\Phi_B) - S_3(\Phi_C)}{T}\,,
\ee
which is the difference between the on-shell action of the bounce solution and that of the false vacuum. Our results show that \eqref{eq:bounce.action} is enhanced near the critical temperature, consistently with expectations from the thin-wall approximation. As the temperature increases, it decreases monotonically, as shown in figure \ref{bact}. For large enough temperatures, it is quite well approximated by the function $S_{3,B}/T=0.2 g (\bar T-1)^{-3/2}$.

The bubble nucleation rate, in the semiclassical limit, is then approximated by
\be
\Gamma =M_{KK}^4 \left( \frac{S_{3,B} }{2\pi T} \right)^{3/2} e^{-S_{3,B} /T}\,,
\ee
where the prefactor $M_{KK}^4$ is an order-of-magnitude estimate of the actual (and challenging to compute) one. It has been obtained  as the value of $V''(\phi_{f})^2$, where $\phi$ is the canonically normalized field and $\phi_{f}$ is its value in the false vacuum (see e.g.~\cite{Hindmarsh:2020hop} for a review).\footnote{In the supercooled case, the same estimate provides the  prefactor $T^4$ used in \cite{Bigazzi:2020phm}.}

\begin{figure}
\center
\includegraphics[scale=0.7]{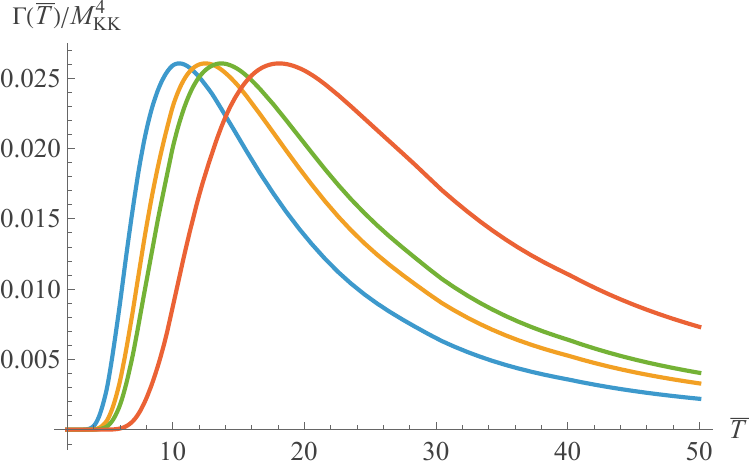}
\caption{Bubble nucleation rates for representative values of $g$ in the ballpark of the standard fits of the holographic model with pure YM or QCD: $g=223$ (blue line), $g=300$ (orange line), $g=350$ (green line), $g=558$ (red line).}
 \label{rates}
\end{figure}

Plots of the nucleation rate for different values of $g$ are shown in figure \ref{rates}.\footnote{\label{footpar} We have chosen $g=\lambda N^2={\cal O}(100)$ since this is  the typical order of magnitude  once  we fix $N=3$ and we fit the holographic model with pure Yang-Mills (YM) or QCD. For instance, $\lambda = 33.26$, $M_{KK} = 949\,\rm{MeV}$ is the Sakai-Sugimoto mesonic  fit \cite{Sakai:2004cn}; $\lambda = 24.88$, $M_{KK} = 790\,\rm{MeV}$ is the Hata-Sakai-Sugimoto baryonic  fit \cite{Hata:2007mb}; and $\lambda = 38.76$, $M_{KK} = 785\,\rm{MeV}$ is the quantum-corrected  baryonic  fit proposed in \cite{Bartolini:2016dbk}. We can also consider a purely Yang-Mills fit, which in the present case is more appropriate. For instance, we can fit the topological susceptibility and the string tension of the holographic model (see e.g.~\cite{Bigazzi:2015bna} for their expressions at generic values of the Yang-Mills $\theta$ angle) with those of (large $N$ and/or $N=3$) Yang-Mills. Using, e.g.,~the lattice results reported in \cite{Bonanno:2025eeb}, we get $\lambda=62$, $M_{KK}=555\,\rm{MeV}$. } 
We notice that the amplitude at the peak does not vary with $g$. Instead, the peak temperature increases as the coupling $g$ increases. Its behavior  is quite well approximated by the function $\bar T_{peak}=1+(2/3)^{2/3}(0.2 g)^{2/3}$.
\subsection{Transition parameters}
\label{sec:transparcdc}
Various transition parameters relevant for gravitational wave phenomenology, like the bubble nucleation temperature $T_n$ and the percolation temperature $T_p$, can be determined once the typical evolution time scale of the whole system is specified. In cosmological contexts, this is provided by the Hubble time. In the case of neutron star mergers, as observed in \cite{Blas:2022xco}, the typical time scale for the overheated regions in which bubbles can nucleate is $\tau\approx 1 \rm{ms}$. The corresponding energy scale
\be
Y=\tau^{-1}\approx 2\cdot 10^{-21} \rm{GeV}
\label{eq:Yvalue}
\ee
plays the role that in the cosmological phase transitions is played by the Hubble rate $H$.

Since, for the moment, we are not turning on any baryon chemical potential in the WSS model, we cannot directly fit it with any high-density QCD phase which could be relevant for neutron star physics. Just to give an idea of the possible values of the transition parameters, let us nevertheless take $Y$ in \eqref{eq:Yvalue} as the relevant scale.

The nucleation temperature is the one at which there is order one bubble nucleated per unit time and volume, so that 
\be
\Gamma[T_n] = Y^4\,.
\ee 
Using the typical fits of the holographic models to pure Yang Mills and QCD (see footnote \ref{footpar}) that give $M_{KK}\in[0.5,1]\ \rm{GeV}$ and $g\in [200, 600]$, we find that its ratio with the critical temperature approximately lies in the interval $\bar T_n\in [1.3,1.7]$ and increases with $g$.

The percolation temperature $T_p$ is defined as the temperature at which the bubbles are nucleated at a sufficiently high rate as to be able to percolate a significant fraction of the space. Following \cite{Blas:2022xco}, it is the solution of
\be
\left(\frac{S_{3,B}}{T}\right)^{-4} \Gamma |_{T_p}= \frac{Y^4}{8\pi v^3}\,,
\label{percol}
\ee
where $v$ is the bubble wall velocity, to be determined from the out-of-equilibrium dynamics. A rough estimate of the velocity will be given in the next subsection with the result shown in eq.~\eqref{eq:vappro)}. Using $v\approx 0.1$ as in \cite{Blas:2022xco}, we find that for the above-mentioned parameter values, the ratio of the percolation temperature to the critical temperature approximately lies in an interval $\bar T_p\in [1.4,1.7]$ and increases with $g$.

The relative transition rate parameter, which measures how rapidly the nucleation rate changes in time and is one of the crucial quantities for the gravitational wave spectrum\footnote{In the cosmological case, it is $\beta/H$.} can be estimated to be
\be
\frac{\beta}{Y} \approx \frac{S_{3,B}}{T}\bigg|_{T_p}\,.
\ee
It turns out to be around $170$ for typical values of the model parameter fits. 
Samples of the relevant transition parameters for different values of $g, M_{KK}$ (and using eq.~\eqref{eq:vappro)} for the bubble wall velocity) are reported in table \ref{tab3}.
\begin{table}[h] 
\centering
\caption{Examples of confinement to deconfinement transition parameters.}
\label{tab3}
\begin{tabular}{|l|c|c|c|c|}
\hline
 & \textbf{$T_c=M_{KK}/2\pi \; (\rm{MeV})$} & \textbf{$T_n \; (\rm{MeV}) $} & \textbf{$T_p \; (\rm{MeV})$} & \textbf{$\beta/Y$} \\ \hline
\textbf{$g = 223$} & $125.796$ & $175.979$ & $183.934$ & $138.455$\\ \hline
\textbf{$g=300$} & $151.115$ & $225.922$ & $235.29$ & $138.499$\\ \hline
\textbf{$g = 350$} & $125.001$ & $193.035$ & $201.348$ & $138.612$\\ \hline
\textbf{$g=558$} & $88.371$ & $153.755$ & $161.582$ & $141.998$\\ \hline
\end{tabular}
\end{table}

The strength parameter that measures the amount of energy absorbed (or released) by the phase transition, relative to the energy density of the surrounding plasma, can be defined as the value at the nucleation temperature of\footnote{Here we adopt the same definition as in \cite{Bigazzi:2020avc}. A slightly different definition can be found, e.g., in \cite{Barni:2025gnm}. Notice that $\rho$ here denotes the energy density.} 
\be
\alpha = \frac{\theta_f - \theta_t}{|\rho_+|}
\label{defalpha}
\ee
where $f/t$ stands for false/true vacuum, 
\be
\theta= \frac{\rho-3p}{4}\,,
\ee
is the trace of the stress energy momentum tensor and $\rho_+$ is the energy density ahead of the wall. 
The sign of $\alpha$ determines whether the transition is a direct (i.e.~supercooled) or inverse (i.e.~superheated) one. In the first case, $\alpha>0$ and the transition releases energy into the environment; in the second case, $\alpha<0$ and the transition absorbs energy from the environment \cite{Barni:2024lkj,Barni:2025gnm}. 

\begin{figure}
\center
\includegraphics[scale=0.7]{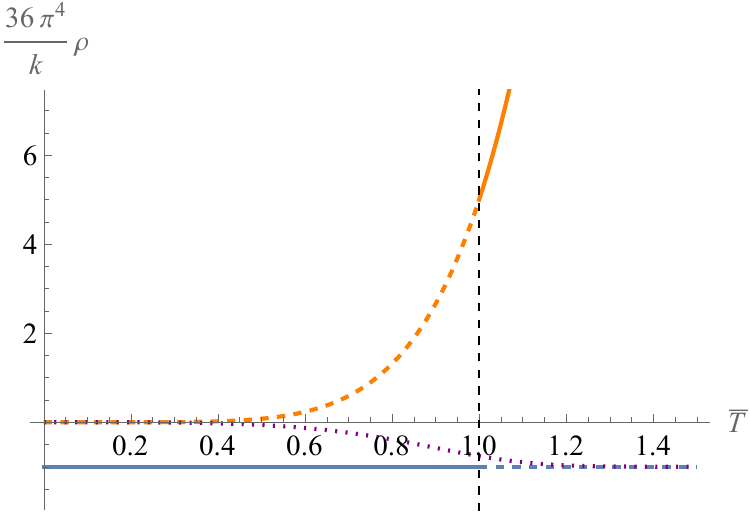}
\caption{Energy density as a function of $\bar T$ in the holographic unflavored model. Solid (resp.~dashed) lines correspond to stable (resp.~metastable) phases. In orange (resp.~blue) the deconfined (resp.~confined) branch. The dotted purple line is just sketched and corresponds to the unstable branch; its actual slope can be computed following \cite{Berenguer:2026ftk}. }
 \label{fig:CdCEnergy}
\end{figure}

Using standard thermodynamic relations, in the present case we have
\be
\rho_+=\rho_f=\rho_C= k V_C = -\frac{k}{36\pi^4}\,,\quad \rho_t=\rho_D=-5k V_D = \frac{5k}{36\pi^4}{\bar T}^6\,,
\label{densitiescdc}
\ee
where
\be
k=\frac{4\pi^2}{3^5}\lambda N^2 M_{KK}^4\,.
\ee
A plot of the energy as a function of the temperature is provided in figure \ref{fig:CdCEnergy}.
Moreover,  we have
\be
p_f= p_C = -k V_C = \frac{k}{36\pi^4}\,,\quad p_t=p_D= -k V_D = \frac{k}{36\pi^4}{\bar T}^6\,.
\label{pressurescdc}
\ee
As a result the strength parameter is given by
\be
\alpha=\frac{2V_C+V_D}{2|V_C|} = -\frac{2+{\bar T}^6}{2}\,.
\ee
For a nucleation temperature $T_n\in [1.3,1.7] T_c$, we get $\alpha(T_n)\in [-13,-3.4]$.

\subsection{An estimate of the bubble wall velocity}
The deconfinement transition we are focusing on is characterized by an extremely large jump in the entropy (or enthalpy) density between the two phases. In particular 
\be
\delta\equiv \frac{w_f}{w_t}=\frac{s_f}{s_t}=0\,,
\ee
to leading order in the large $N$ and large $\lambda$ limits, since the entropy density of the false vacuum (the confined phase) is zero, as the pressure does not depend on the temperature. 

In this case, deflagrations are argued to be highly favored \cite{Sanchez-Garitaonandia:2023zqz}. A deflagration corresponds to a bubble wall that propagates slower than the speed of sound of the plasma behind it, leaving the latter effectively unperturbed. In the inverse deconfinement phase transition of the WSS model, the large jump in entropy, and correspondingly in degrees of freedom, implies a substantial transfer of energy to the deconfined phase behind the wall. This leads to large friction effects and makes deflagrations more likely to occur during this transition. An estimate of the bubble wall velocity $v$, following the suggestions in \cite{Sanchez-Garitaonandia:2023zqz}, just allows to conclude that $v\sim\delta$ is indeed highly suppressed in our case. 

Let us thus  provide an estimate of the velocity that might be sensible in the thin-wall approximation, and thus for temperatures quite close to $T_c$. In the wall frame, the velocities $v_{-}$ and $v_+$ inside and outside the bubble can be given in terms of thermodynamic variables as follows
\be
v_+ v_- = \frac{p_+ - p_-}{\rho_+-\rho_-}\,,\quad \frac{v_+}{v_-}=\frac{\rho_-+p_+}{\rho_++p_-}\,.
\ee
If we now trade $+/-$ with $f/t$ (which is a reasonable approximation only for small enough bubble wall velocities), assuming that on the left and on the right of the wall the fluid has approximately the same temperature, and we use the results in (\ref{densitiescdc}) and (\ref{pressurescdc}), we find that
\be
v_+ v_- = \frac{v_-}{v_+} = \frac{\bar T^6-1}{1+5\bar T^6}\,,
\ee
which implies that
\be
v_-= \frac{\bar T^6-1}{1+5\bar T^6}\,.
\ee
This is related to the corresponding velocity $\tilde v_{-}$ in the lab frame by
\be
|v_-| = |\frac{\tilde v_--v}{1-\tilde v_-\, v}|\,.
\ee
For deflagrations $\tilde v_-=0$ and thus
\be
\label{eq:vappro)}
|v| = |v_-| = \frac{\bar T^6-1}{1+5\bar T^6} =\bar T -1 -\frac52(\bar T - 1)^2+{\cal O}((\bar T-1)^3)\,,
\ee
which goes to zero in the $\bar T\rightarrow 1$ limit, consistently with our assumptions.\footnote{Extrapolating the result at any $T>T_c$, it would anyway provide a velocity ranging in the $[0,0.2]$ interval.}

\section{Chiral symmetry restoration transition}
\label{eq:DBI}

\subsection{Setup}

In this section, we consider  the deconfined phase of the WSS model, which arises for temperatures $T>M_{KK}/2\pi$. Above this threshold,  a first-order chiral phase transition may occur. The direct transition from the chirally restored phase to the chirally  broken one has been studied in \cite{Bigazzi:2020phm}. 

Here, we focus instead on the reverse process, which may occur upon heating the system. For temperatures above the critical value for chiral symmetry breaking, the metastable chiral symmetry broken phase (the false vacuum) can decay through bubble nucleation into the energetically favored chiral symmetry restored phase (the true vacuum). 

In the WSS model, this process is described by a bounce solution interpolating between a connected U-shaped flavor brane configuration and  a disconnected embedding in which the branes fall orthogonally into the horizon of the black hole background \eqref{bhme} dual to the deconfined phase of the model. Defining\footnote{Remember, from section \ref{S:WSS}, that $L$ is the asymptotic separation  between the flavor branes.}
\be
\tilde L \equiv \frac{4\pi}{3} L\,T\,,
\ee
for such a solution to exist, the temperature must not only exceed the critical value $\tilde{L}_{c}\approx 0.6444$, but also satisfy $\tilde{L}<0.7016$, above which connected flavor brane configurations cease to exist \cite{Aharony:2006da,Bigazzi:2020phm}. Thus, in what follows, we restrict to 
\be
\label{eq:range}
0.6444<\tilde{L}<0.7016\,,\qquad \text{or}\qquad 1<\ \frac{T}{T^\chi_c}<1.089\,,
\ee
where $T^\chi_{c}\approx 0.1538L^{-1}$, $\tilde{L}_{c}\equiv \frac{4\pi}{3} L\,T^\chi_{c}\approx 0.6444$.\footnote{If the external dynamics, e.g., a neutron star merger, drives the production of superheated regions to temperatures $T/T^\chi_c>1.089$, there is no metastable connected phase and the transition proceeds instantaneously without  bubble nucleation.}

The Dirac-Born-Infeld (DBI) action describing the profile $x_4 = x_4(u)$ of a $D8$ flavor brane extended along the $t,x^i,u,S^4$ directions in the black-hole background (\ref{bhme}) reads
\be
S_{DBI}=\frac{T_8}{g_s}\int d^9x \left(\frac{u}{R}\right)^{-3/2}u^4 \sqrt{1 + f_T(u) \left(\frac{u}{R}\right)^3 (\partial_u x_4)^2}\,,
\label{SDBI1}
\ee
where $T_8=(2\pi)^{-8}l_s^{-9}$. After integration along the $t,x^i,S^4$ directions, the action can be rewritten as
\be
S_{DBI}= K\int y^{5/2}\sqrt{1+y^3f_T(\partial_yx)^2}dy\,,
\label{redefDBI}
\ee
where 
\be
\label{eq:defKappa}
K=(2\pi)^{-8}l_s^{-9}g_s^{-1}\frac{V_3}{T} V_{S^4}R^{3/2}u_T^{7/2} = \frac{8\pi^2}{3^8} \frac{N V_3 T^6 \lambda^3}{M_{KK}^3}\,,
\ee
with $V_3=\int d^3x$ being the infinite spatial three-dimensional volume and $V_{S^4}=8\pi^2/3$ being the volume of the internal four-sphere. In (\ref{redefDBI}) we have introduced dimensionless variables $x,y$ defined as
\begin{equation}
x_4 = x \,  u_T^{-1/2} R^{3/2} = x \frac{3}{4\pi T} \ , \qquad u = y\,u_T \,,
\label{redef}
\end{equation}
such that
\begin{equation}
f_T(u) \equiv f_T = 1 - y^{-3}\ .
\label{redef2}
\end{equation}

The Euler-Lagrange equation for $x_4(u)$ has to be solved requiring that $x_4(\infty)=\pm L/2$, for $L$ centered around $x_4=0$. Here, the two signs correspond to the brane-antibrane asymptotic conditions. Equivalently, for the dimensionless embedding $x(y)$, we have to require that $x(\infty)=\pm {\tilde L}/2$. The Euler-Lagrange equation, then, admits two types of solutions.
One is the disconnected configuration 
\be
\label{eq:disc}
x_d=\pm \frac{\tilde{L}}{2}\,,
\ee
for which $\partial_yx=0$
and the corresponding on-shell action is
\be\label{eq:Sd}
S_{DBI}[x_d]= 2K\int_1^{\infty} y^{5/2}dy\,.
\ee
The factor of 2 accounts for the brane-antibrane pair. Notice that this action is divergent. In what follows, we will always consider differences in on-shell actions such that divergent terms cancel.

\begin{figure}
\center
\includegraphics[scale=0.5]{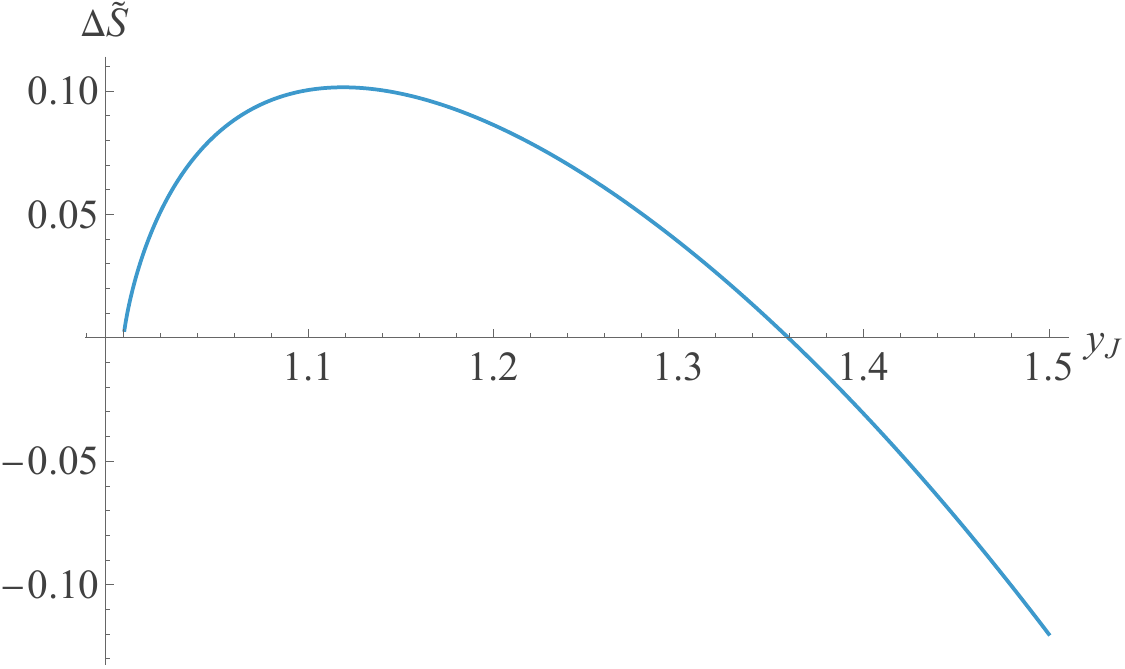}
\caption{The difference between the normalized on-shell disconnected and connected actions as a function of $y_J$.}
 \label{fig:deltaStilde}
\end{figure}

The other possible solution is the U-shaped connected configuration where the brane-antibrane branches meet at a turning point $y_J$ for which  
$(\partial_y x)\big|_{y_J}=\infty$\,. 
As it was shown in \cite{Bigazzi:2020phm}, such a solution is extremely well approximated by the variational ansatz\footnote{We refrain from adding a subscript $x_{var}(y)$ to keep the notation simple.}   
\be\label{eq:ansatz}
x(y)=\pm \frac{\tilde{L}}{2} {\rm tanh}\left(\frac{\sqrt{y-y_J}}{\sqrt{B}}\right)\,.
\ee
The parameters $y_J$ and $B$ are variational constants that take values $1\leq y_J <\infty$ and $B>0$. This variational profile interpolates between the connected configurations and the disconnected profile for $B\rightarrow 0$. For a particular value of the brane-antibrane insertion given by $\tilde{L}$, the values of $y_J$ and $B$ that best approximate the connected solution can be derived by minimizing 
\be
\label{eq:DeltaS}
\Delta S= S_{DBI,var}-S_{DBI}[x_d]\,,
\ee
where $S_{DBI}[x_d]$ is the on-shell action given  in \eqref{eq:Sd} and $S_{DBI,var}$ is the action given by 
\be
S_{DBI,var}=2K\int_{y_J}^{\infty} \sqrt{1+f_Ty^3(\partial_yx)^2}\,,
\ee
 with the ansatz \eqref{eq:ansatz}. A sample of values of $\tilde{L}$, $y_J$, and $B$ in the range \eqref{eq:range} considered in this work, obtained by numerically minimizing  \eqref{eq:DeltaS}, can be found in table \ref{T:LyjB}. 
 Notice that for all these values, the on-shell action difference $\Delta S[x]=S_{DBI,var}[x]-S_{DBI}[x_d]>0$ and it is decreasing with $y_J$ (for smaller values of $y_J$ it is increasing: this corresponds to the unstable branch). Since $T \Delta S = \Delta \cal{F}$, where $\Delta\cal{F}$ is the free energy difference, the disconnected brane configuration is thus always preferred over this temperature range. Using the variational ansatz, the transition occurs for $\tilde{L}=0.6442$, $y_J=1.360$, and $B=0.2723$ after which $\Delta S[x]<0$ signaling that the connected configuration is energetically preferred. See again \cite{Bigazzi:2020phm} for details of the derivation and figure \ref{fig:deltaStilde} for a plot of the normalized on-shell action difference $\Delta \tilde S\equiv K^{-1} \Delta S$ as a function of $y_J$.

\begin{table}[h]
\centering
\caption{Sample of values of $\tilde{L}$, $y_{J}$ and $B$ that minimize $\Delta S[x]$ with the variational ansatz \eqref{eq:ansatz}.}
\begin{tabular}{|l|l|l|l|}
\hline
\textbf{$\tilde{L}$}
& \textbf{$y_J$} & \textbf{$B$}& \textbf{$\Delta S[x]/K$}  \\ \hline
$0.645 $ & $1.357$ & $0.2714$ & $0.00204$
\\ \hline
$0.650 $ & $1.339$ & $0.2656$ & $0.01466$
\\ \hline
$0.660 $ & $1.303$ & $0.2540$ & $0.03743$
\\ \hline
$0.670 $ & $1.267$ & $0.2421$ & $0.05716$
\\ \hline
$0.680 $ & $1.231$ & $0.2294$ & $0.07410$
\\ \hline
$0.690 $ & $1.193$ & $0.2150$ & $0.08841$
\\ \hline
$0.700 $ & $1.144$ & $0.1943$ & $0.09994$
\\ \hline
\end{tabular}
\label{T:LyjB}
\end{table}
\subsection{The $O(3)$ symmetric bubble}
To compute the parameters relevant for the nucleation of bubbles of the true chirally symmetric vacuum inside a chirally broken medium, we need to determine the bounce solution of the equations of motion. This solution interpolates between the connected false-vacuum configuration and the disconnected true-vacuum brane configuration. 
The original brane embedding $x(y)$ describes a homogeneous vacuum configuration, in which every spatial point in the dual theory is in the same phase. 
A bubble, however, is a localized object: near its center,  the system is expected to be close to the true vacuum, while at large distances it approaches the false vacuum. We must therefore allow for inhomogeneous configurations.

Since the decay takes place at a finite temperature, the Euclidean time direction is compactified.  The dominant bounce is expected to be static along the Euclidean time, reducing the problem to three spatial dimensions \cite{Linde:1980tt}. Thus, the relevant configuration is $O(3)$ symmetric, and its dependence on the spatial coordinates reduces to a dependence on the radial variable $\rho=\sqrt{x^i x^i}$, which in the following will be traded by its dimensionless version 
\be
\label{eq:defsigma}
\sigma=\frac{4\pi T}{3}\rho\,.
\ee
We therefore  promote the brane embedding $x(y)$ to be a function of both the holographic coordinate $y$ and the radial coordinate $\sigma$,  $x(y,\sigma)$. The  corresponding DBI action  is given by
\be
\label{eq:defiS}
S_{DBI}\equiv \frac{S_{3}}{T}= \tilde{K}\tilde{S}\,,
\ee
with
\be
\label{eq:SDBI3}
\tilde{S}= 2\int d\sigma \sigma^2\int y^{5/2}\sqrt{\Delta}dy\,,
\ee
and
\be
\label{eq:Delta}
\Delta=1+(y^3-1)(\partial_y x)^2+(\partial_{\sigma}x)^2\,,
\ee
as well as (using \eqref{eq:defKappa} and \eqref{eq:defsigma})
\be
\label{eq:deftildeK}
\tilde{K}=\frac{N T^3 \lambda^3}{486 M_{KK}^3}\,.
\ee

The corresponding Euler-Lagrange equations are 
\be\label{eq:EL.bounce}
\partial_y\pi^y+\partial_{\sigma}\pi^\sigma=0\,,
\ee
where 
\bea
\pi^y=\frac{1}{\sqrt{\Delta}}\sigma^2 y^{5/2}(y^3-1)(\partial_y x)\,,\qquad \pi^{\sigma}=\frac{1}{\sqrt{\Delta}}\sigma^2 y^{5/2}(\partial_{\sigma}x)\,.
\eea
It is important to notice that these equations may admit disconnected brane-antibrane configurations that end at the horizon non-orthogonally, that is, satisfying  $0<(\partial_y x)<\infty$ at $y=1$. This can be understood as follows.
Notice that for the previous simple  brane embedding $x=x(y)$, $\pi^{\sigma}=0$ and the EL  equation simplifies to 
\be
\partial_y \left(\frac{y^{5/2}(y^3-1)(\partial_y x)}{\sqrt{1+(y^3-1)(\partial_yx)^2}}\right)=0\qquad \rightarrow \qquad (\partial_y x)^2=\frac{C^2}{(y^3-1)(y^5(y^3-1)-C^2)}\,,
\ee
where $C$ is a constant. If $C\neq 0$, close to the horizon $y\rightarrow 1$ we have 
\be
(\partial_{y}x)\sim \frac{C}{(y^3-1)}\rightarrow \infty\,.
\ee
For $C=0$, then the only solution is 
\be
(\partial_y x)=0\qquad \rightarrow \qquad x={\rm constant}\,,
\ee
corresponding to disconnected branes entering the horizon orthogonally. 
For the general bubble embedding $x=x(y,\sigma)$, $\pi^y$ is generically not conserved, and the previous argument forcing $C=0$ for disconnected configurations does not apply. Thus, disconnected brane configurations that enter the horizon non-orthogonally   $0<(\partial_y x)<\infty$ are in principle allowed solutions.

As a consequence, we expect the bounce solution to interpolate between 
a disconnected brane-antibrane pair
at the center of the bubble, $\sigma=0$, and the connected false-vacuum configuration at asymptotic infinity,  $\sigma\rightarrow\infty$. Along this interpolation, the solution may pass through several disconnected but ``non-vertical'' (i.e.~not orthogonal to the horizon) configurations before becoming connected and eventually approaching the false vacuum profile. 

In what follows, we proceed in two steps. First, for illustrative purposes, we construct an approximate bounce solution in which the configuration has exactly the true-vacuum disconnected profile inside the bubble, while outside it interpolates between connected configurations. Then, we consider a setup more closely aligned  with the expectation described above, in which the true-vacuum disconnected vertical configuration 
is never reached.
In both cases, we do not solve the exact partial differential equations  \eqref{eq:EL.bounce} but rely on a variational ansatz of increasing accuracy.  We compare the two solutions at the end of this section and show that their orders of magnitude are similar.

\subsubsection{An approximate solution}

To compute the $O(3)$ bounce solution, we have introduced a radial dependence of the brane embedding function $x(y,\sigma)$. In the variational ansatz \eqref{eq:ansatz}, this can be implemented by promoting the  variational parameters to be $\sigma$-dependent $y_J\rightarrow y_J(\sigma)$ and $B\rightarrow B(\sigma)$ such that  
\begin{equation}
x(y,\s)=\pm\frac{\tilde L}{2} \tanh\left(\frac{\sqrt{y-y_J(\s)}}{\sqrt{B(\s)}}\right) \,.
\label{vari_profile3}
\end{equation}

In principle, by inserting this ansatz into the $O(3)$ DBI action \eqref{eq:SDBI3}, it is possible to derive two second-order coupled differential equations for $y_{J}(\sigma)$ and $B(\sigma)$ and solve them independently. However, as in \cite{Bigazzi:2020phm}, we further assume
 $y_{J}(\sigma)$ and $B(\sigma)$ to depend on a single function $\alpha(\sigma)$. By simply inverting the ansatz of \cite{Bigazzi:2020phm}, we require
\begin{eqnarray}
y_J(\s)&=& 1 -  (1-y_{J,f})\alpha(\s) \ ,\nonumber\\
B(\s)&=&B_{f}\,\alpha(\s) \ ,
\label{y0Balpha}
\end{eqnarray}
where the $f$ labels stands for false vacuum. This ansatz ensures that for $\alpha(\sigma)=1$, the configuration is in the false vacuum and for $\alpha(\sigma)=0$ it corresponds to the true vacuum (the disconnected configuration), for which  $B=0$ (and correspondingly $y_J=1$).  

As anticipated, here we look for an approximate bounce solution that interpolates between connected configurations  outside the bubble  $\sigma>\sigma_0$, that is we require 
\be
\alpha(\sigma_0) =0\,,\qquad \alpha'(\sigma_0)=0\,,\qquad \alpha(\infty)=1\,\qquad \text{for}\qquad \sigma\ge\sigma_0\,,
\label{alpha1}
\ee
and it is smoothly connected to the constant solution inside the bubble
\be
\alpha(\sigma)=0\, \qquad \text{for}\qquad  \sigma<\sigma_0\,,
\label{alpha2}
\ee 
which describes the true vacuum. The parameter $\sigma_0$ needs to be determined via a shooting procedure for the solution with these boundary conditions to exist. 
 Although the variational approach does not provide a solution for $\alpha(\sigma)$ which satisfies both (\ref{alpha1}) and (\ref{alpha2}), we know that the disconnected configuration is an exact solution to the complete problem. Hence 
we use this exact solution to satisfy (\ref{alpha2}) and the variational approach to satisfy (\ref{alpha1}).

For any $\tilde{L}$, the variational part of the bounce solution can then be found by extremizing the $O(3)$ DBI action \eqref{eq:SDBI3}. For numerical convenience, we work with the subtracted Euclidean DBI action  $\Delta \tilde{S}= \tilde{S}-\tilde{S}[x_d]$ instead, where $\tilde{S}[x_d]$ is simply 
\be
\label{eq:S.tv}
\tilde{S}[x_d]=2\int_0^{\infty} d\sigma \sigma^2\int_{1}^{\infty}y^{5/2}dy\,.
\ee
The corresponding Lagrangian in $\Delta \tilde{S}=\int  d\sigma{\cal L}$ can be written as
\be
{\cal L}= \int_0^{\infty}Fdz+G\,,
\ee
where
\bea
F&=&2\sigma^2 B(\sigma)(B(\sigma)z+y_{J}(\sigma))^{5/2}(\sqrt{\Delta}-1)\,,\\
G&=&-\frac{4}{7}\sigma^2(y_{J}(\sigma)^{7/2}-1)\,,
\eea
with $\Delta$ given in \eqref{eq:Delta} on the variational ansatz \eqref{vari_profile3} and \eqref{y0Balpha}.
We have also defined 
\be
\label{eq:change.z}
z=\frac{y-y_J(\sigma)}{B(\sigma)}\,,
\ee
for the endpoints of integration to be finite variables.

\begin{figure}[htb]
\center
\includegraphics[scale=0.6]{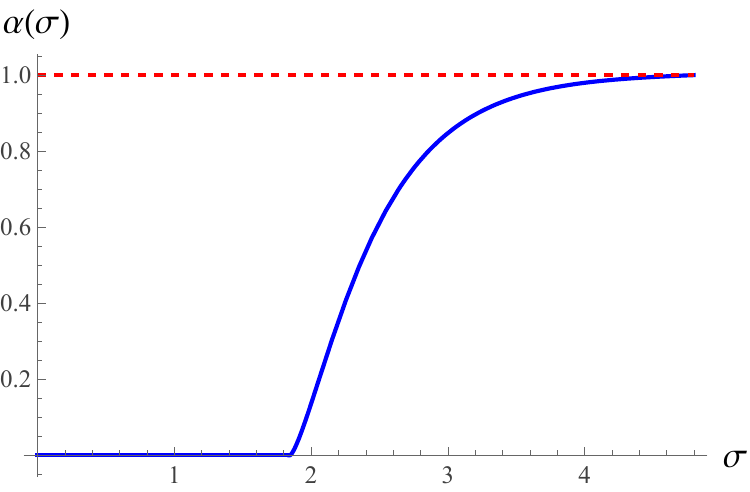}
\includegraphics[scale=0.6]{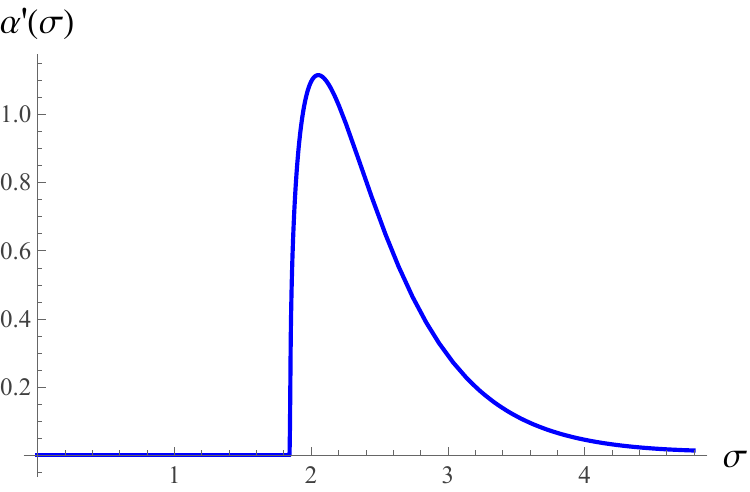}
\caption{The solution for $\alpha(\s)$ and its derivative at $\tilde L=0.685$.}\label{DBIbouncef}
\end{figure}

The resulting second-order differential equation for $\alpha(\sigma)$ is
\be
\alpha''(\sigma)=\left(\int_0^\infty Jdz\right)^{-1}\left( -\int_0^{\infty}Hdz+\int_{0}^{\infty}\left(\frac{\partial F}{\partial \alpha(\sigma)}dz\right)+\left(\frac{\partial G}{\partial \alpha(\sigma)}\right)\right)\,,
\ee
where we have also defined 
\be
\frac{d}{d\sigma}\left(\frac{\partial {\cal L}}{\partial \alpha'(\sigma)}\right)=\int_0^{\infty}Hdz+\alpha''(\sigma)\int_0^{\infty}Jdz\,,
\ee
and it is possible to find a solution numerically, subject to the boundary conditions prescribed by \eqref{alpha1}. It turns out that if $\sigma_0$ is chosen to be too big, then the solution for $\alpha(\sigma)$ becomes $\alpha(\sigma)>1$ asymptotically (it overshoots). If $\sigma_0$ is too small, then $\alpha(\sigma)$ never reaches the condition $\alpha=1$ (it undershoots). The solution can then be found by adjusting the parameter $\sigma_0$ such that the boundary conditions \eqref{alpha1} are just met. 
 A sample of the solution $\alpha(\sigma)$ is provided in figure \ref{DBIbouncef}.

\begin{figure}[htb]
\center
\includegraphics[scale=0.7]{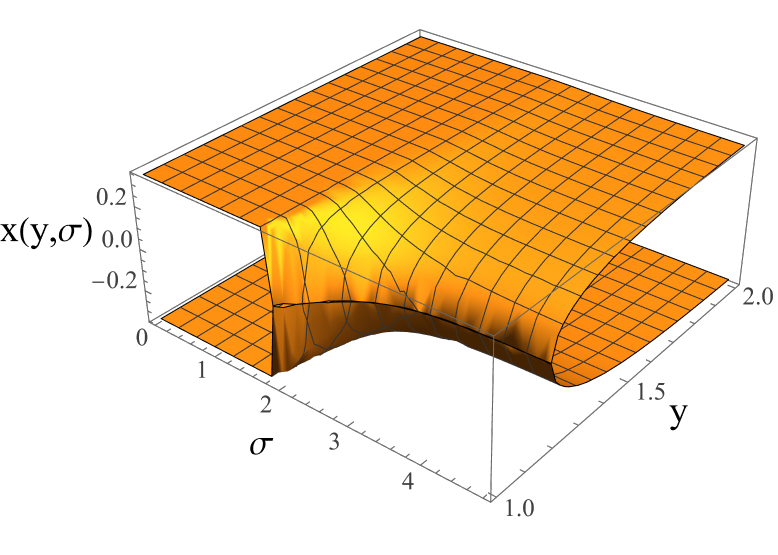}
\caption{The bubble solution for $\tilde L=0.685$.}
\label{fig:3dbounce}
\end{figure}

The corresponding bounce solution $x_B(y,\sigma)$ that can be obtained by inserting $\alpha(\sigma)$ into \eqref{y0Balpha} and  \eqref{vari_profile3} for the entire profile is shown in figure \ref{fig:3dbounce}. 

\begin{figure}[htb]
\center
\includegraphics[scale=0.7]{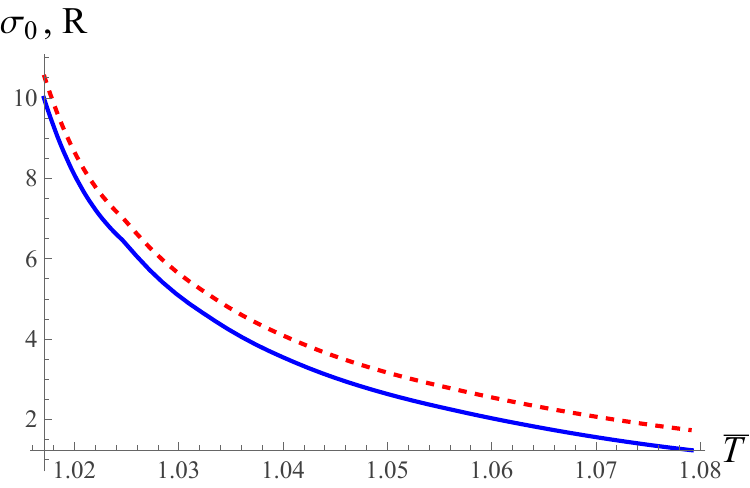}
\caption{Comparison plot between $\sigma_0$ trend (blue thick line) and bubble radius $R$ trend (red dashed line), both as functions of $\bar T$.}
\label{fig:r0_rad_confr}
\end{figure}

The values of $\sigma_0$ as a function of 
\be
\bar T \equiv \frac{T}{T^\chi_c}\,,
\ee
are reported in figure \ref{fig:r0_rad_confr}. 
These approximate very well the dimensionless radius of the bubble $R$,\footnote{Not to be confused with $R$, the scale of the deconfined background defined previously around \eqref{bhme}.} defined here as the value of $\sigma$ for which $\alpha$ is halfway between the true and false vacuum
 \be
 \alpha(R)=\frac{1}{2}\,,
 \ee
 and reported in the same figure \ref{fig:r0_rad_confr}. Thus, we can equivalently take $\sigma_0$ to be the definition of the (dimensionless) bubble radius.
 Notice  that the farther we are from the critical value $\bar T=1$ or $\tilde{L}=0.6442$, the smaller the bubble radius. For temperatures close to the critical value, the bubble radius increases and eventually diverges, as also observed in the direct phase transition \cite{Bigazzi:2020phm} and expected from the thin-wall approximation. Moreover, we notice that closer to the chiral symmetry restoration transition, the profile of $\alpha(\sigma)$ is sharper, with the bubble radius $R$ more closely  related to $\sigma_0$. We interpret this behavior as expected, given the validity of the thin-wall approximation near criticality.   

\begin{figure}[htb]
\center
\includegraphics[scale=0.7]{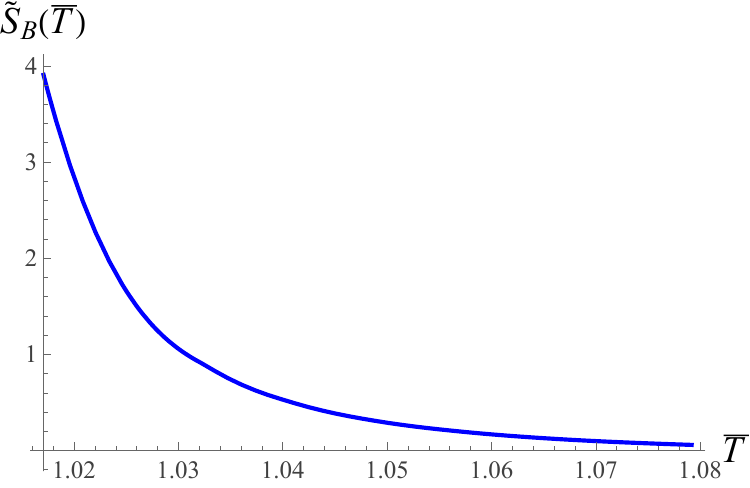}
\caption{The on-shell bounce action $\tilde{S}_B[x_B]$ as a function of $\bar T$.}
\label{fig:S}
\end{figure}

Once the profile of $\alpha(\sigma)$ for any given $\tilde{L}$ is known, it is also possible to compute the on-shell bounce action
\be
\tilde{S}_B[x_B]=\tilde{S}[x_B]-\tilde{S}[x_{f}]\,.
\ee
Here  $\tilde{S}[x_B]$ is the $O(3)$ DBI action \eqref{eq:SDBI3} evaluated on the bounce solution and  $\tilde{S}[x_{f}]$ is   evaluated on the connected configuration describing the false-vacuum
\be
\label{eq:false.vacuum}
x_{f}=\pm\frac{\tilde L}{2} \tanh\left(\frac{\sqrt{y-y_{J,f}}}{\sqrt{B_{f}}}\right)\,,
\ee
where samples of the values $y_{J,f}$ and $B_{f}$ can be inferred from table \ref{T:LyjB}.
More explicitly, the  bounce action can be written as
\be
\tilde{ S}_B[x_B] = 2\int_0^{\infty}d\sigma \sigma^2\left [\int_{y_J(\sigma)}^{y_{J,f}}dy\, y^{5/2}\sqrt{\Delta[x_B]}+\int_{y_{J,f}}^{\infty}dy\, y^{5/2}\left(\sqrt{\Delta[x_B]}-\sqrt{\Delta[x_{f}]}\right) \right]\,,
\label{bacti}
\ee
where $\Delta$ has been defined in \eqref{eq:Delta} and $\Delta[x]$ signifies it is evaluated on the specific solution $x$. The values of the bounce action as a function of $\bar T$ are reported in figure \ref{fig:S}. 
As for the supercooled case studied in \cite{Bigazzi:2020phm}, the bounce action diverges near criticality and approaches zero at higher temperatures, following a behavior similar to that of the bubble radius. 

We believe that this solution  provides a first reasonable approximation to the actual bounce solution. The reason for treating this as an approximation is that the solution does not capture the typical behavior we expect at the bubble's center. As we already discussed, inside the bubble, the bounce solution usually does not correspond to the true vacuum, whereas in this case it does. In the next subsection, we improve on our approximation by allowing non-orthogonal disconnected configurations inside the bubble.


\subsubsection{A fully variational solution}

If the solution inside the bubble ends at the horizon, but it is generically different from the true vacuum, we might have to admit that inside the bubble, the disconnected branes fall into the horizon non-orthogonally. This could be realized if we allow $y_J(\sigma)$ in \eqref{vari_profile3} to be less than one, such that the tip of the U-shaped configuration would lie beyond the horizon and the branes are effectively disconnected but non-vertical. 

To accommodate  this case, we should modify the variational ansatz \eqref{y0Balpha}, for instance, as
\bea\label{eq:var.ansatz}
y_{J}(\sigma)=\alpha(\sigma)y_{J,f}\,,\qquad
B(\sigma)=\alpha(\sigma)B_{f}\,,
\eea
with boundary conditions
\bea
&&\alpha(+\infty)=1\,,\qquad y_{J}(+\infty)=y_{J,f}\,,\qquad B(+\infty)=B_{f}\,,\\
&&\alpha(0)=\alpha_0\,,\qquad \alpha'(0)=0\,,\qquad y_{J}(0)=\alpha_0\,y_{J,f}\,.
\eea
With this ansatz, the bounce solution reaches the false vacuum configuration \eqref{eq:false.vacuum} asymptotically for $\sigma=+\infty$,
 while at the center of the bubble,  $\sigma=0$, it is given by 
 a disconnected configuration if $y_J(0)\le 1$.

The interpolating bounce solution that we seek  will therefore start from 
a certain disconnected 
configuration at $\sigma=0$, will pass through several non-orthogonal disconnected configurations until some value
\be 
\sigma=\sigma_i\,,\quad  \text{ for which} \quad \alpha_i\equiv\alpha(\sigma_i)=\frac{1}{y_{J,f}}\,,
\ee
where it becomes connected as $y_J(\sigma_i)=1$, and then reaches the false vacuum values asymptotically at $\sigma=+\infty$.
To determine such bounce solution, we need to separate the contributions of the DBI action arising from  the disconnected and the connected brane-antibrane configurations
\be
\tilde{S}=\tilde{S}_{dis}+\tilde{S}_{con}\,,
\ee
with
\bea
\tilde{S}_{dis}&=&2\int_{0}^{\sigma_i}d\sigma \sigma^2 \int_{1}^{\infty}y^{5/2}\sqrt{\Delta}dy\,,\\
\tilde{S}_{con}&=&2\int_{\sigma_i}^{\infty}d\sigma \sigma^2 \int_{y_{J}(\sigma)}^{\infty}y^{5/2}\sqrt{\Delta}dy\,,
\ea
where $\Delta$ has been defined in \eqref{eq:Delta}
 and we use  \eqref{vari_profile3} with \eqref{eq:var.ansatz}.

The bounce action that interpolates between the true and  false vacuum profiles is then given by 
\bea
\tilde{S}_B&=&\tilde{S}-\tilde{S}[x_{f}]\,,
\ea
where $\tilde{S}[x_{f}]$ is the $O(3)$ DBI action evaluated on the false vacuum  \eqref{eq:false.vacuum},
\be
\tilde{S}[x_{f}]=2\int^{\infty}_0d\sigma \sigma^2\int_{y_{J,f}}^{\infty}y^{5/2}\sqrt{\Delta[x_{f}]}dy\,.
\ee
To make each integral in $y$ well-defined and finite, it is convenient to split $\tilde{S}_B$ as follows
\be 
\tilde{S}_B =\tilde{S}-\tilde{S}[x_d]-(\tilde{S}[x_{f}]-\tilde{S}[x_d])\,,
\ee
where $\tilde{S}[x_d]$ is the $O(3)$-symmetric DBI action for the true vacuum disconnected configuration  defined in \eqref{eq:S.tv}. 
Therefore we have 
\be
\label{eq:bounce}
\tilde{S}_{B}=\int_{0}^{\infty}{\cal L}\,d\sigma=\int_{0}^{\sigma_i} {\cal L}_{dis}\,d\sigma+\int_{\sigma_i}^{\infty} {\cal L}_{con}\,d\sigma\,,
\ee
with 
\bea
{\cal L}_{dis}&=&\int_{1}^{\infty}F_{dis}dy +C_{dis}\,,\\
{\cal L}_{con}&=&\int_{0}^{\infty}F_{con}dz +G_{con}+C_{con}\,,
\ea
where we have defined
\bea
F_{dis}&=&2\sigma^2 y^{5/2}(\sqrt{\Delta}-1)\,,\\
C_{dis}&=&-2\sigma^2\int_{y_{J,f}}^{\infty}y^{5/2}(\sqrt{\Delta[x_{f}]}-1)+\frac{4}{7}\sigma^2(y_{J,f}^{7/2}-1)\,,
\ea
and
\bea
F_{con}&=&2\sigma^2 B(\sigma)(y_{J}(\sigma)+zB(\sigma))^{5/2}(\sqrt{\Delta}-1)\,,\\
G_{con}&=&-\frac{4}{7}\sigma^2 (y_{J}(\sigma)^{7/2}-y_{J,f}^{7/2})\,,\\
C_{con}&=&-2\sigma^2\int_{y_{J,f}}^{\infty}y^{5/2}(\sqrt{\Delta[x_{f}]}-1)dy\,,
\ea
where 
 in $F_{con}$ we have performed a change of variables \eqref{eq:change.z}
such that the limits of integration are finite. While the $C's$ only depend on $\sigma$ and $y$ or $z$, $G$ also depends on $\alpha(\sigma)$ through    \eqref{eq:var.ansatz}, while $F$ depends on both  $\alpha(\sigma)$ and $\alpha'(\sigma)$ through $(\partial_{\sigma}x)$ in \eqref{eq:Delta}.
   
To find the bounce solution, it is now necessary to derive the Euler-Lagrange equations for $\alpha(\sigma)$. They are given by 
\bea 
\label{eq:in}
\alpha''(\sigma)&=&\left(\int_{1}^{\infty}J_{dis}\,dy\right)^{-1}\left(\int_{1}^{\infty}\left(\frac{\partial F_{dis}}{\partial \alpha(\sigma)}\right)dy-\int_{1}^{\infty}H_{dis}\,dy\right)\quad \text{for}\quad \sigma<\sigma_i\,,\\
\label{eq:out}
\alpha''(\sigma)&=&\left(\int_{0}^{\infty}J_{con}\,dz\right)^{-1}\left(\int_{0}^{\infty}\left(\frac{\partial F_{con}}{\partial \alpha(\sigma)}\right)dz+\left(\frac{\partial G_{con}}{\partial \alpha(\sigma)}\right)-\int_{0}^{\infty}H_{con}\,dz\right)\quad \text{for}\quad \sigma>\sigma_i\,,\nonumber\\
\ea   
 where we have defined 
 \bea 
 \frac{d }{d \sigma}\left(\frac{\partial F_{dis}}{\partial \alpha'(\sigma)}\right)\equiv \int_{1}^{\infty}H_{dis}\,dy+\alpha''(\sigma)\int_{1}^{\infty}J_{dis}\,dy\,,\\
  \frac{d }{d \sigma}\left(\frac{\partial F_{con}}{\partial \alpha'(\sigma)}\right)\equiv \int_{0}^{\infty}H_{con}\,dz+\alpha''(\sigma)\int_{0}^{\infty}J_{con}\,dz\,,
 \ea 
and $H$ and $J$ do not depend on $\alpha''(\sigma)$. Notice that the $C$'s terms do not enter the Euler-Lagrange equations; they will nevertheless determine the value of the on-shell Euclidean bounce action. 

We solve these equations numerically with a shooting method with the initial conditions  given by 
\be 
\alpha(0)=\alpha_0\,,\qquad \alpha'(0)=0\,.
\ee
The shooting parameter $\alpha_0$ now defines the brane configuration at the center of the bubble, which is close to the true vacuum configuration but not necessarily equal. It needs to be adjusted to match the asymptotic value 
\be
\alpha(\infty)=1\,,
\ee
provided that we are careful in using the differential equation \eqref{eq:in} for disconnected configurations up to the point $\sigma_i$ for which $\alpha(\sigma_i)=y_{J,f}^{-1}$, and then continuously switch to \eqref{eq:out} feeding into it as the initial conditions, the last values $\sigma_i$, $\alpha(\sigma_i)$ and $\alpha'(\sigma_i)$.
For a given $\tilde{L}$, if $\alpha_0$ is too small, there will be overshooting and $\alpha(\infty)>1$, while if $\alpha_0$ is too big, $\alpha(\infty)=1$ will never be reached and the solution will start to decrease.

\begin{figure}[htbp]
\center
\includegraphics[scale=0.6]{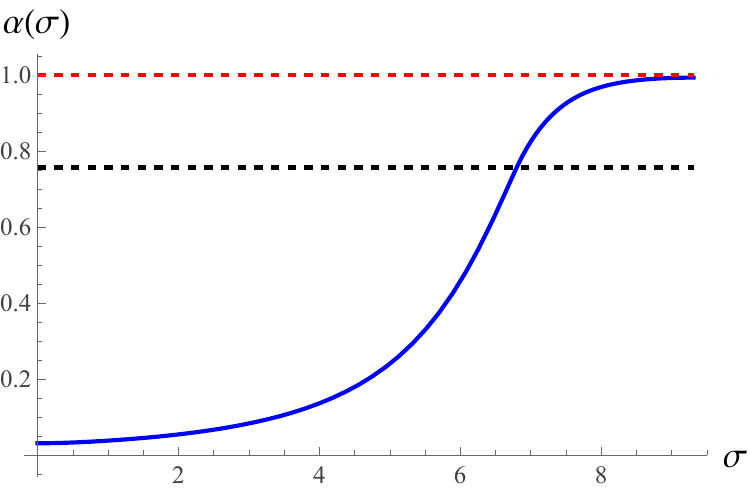}
\includegraphics[scale=0.6]{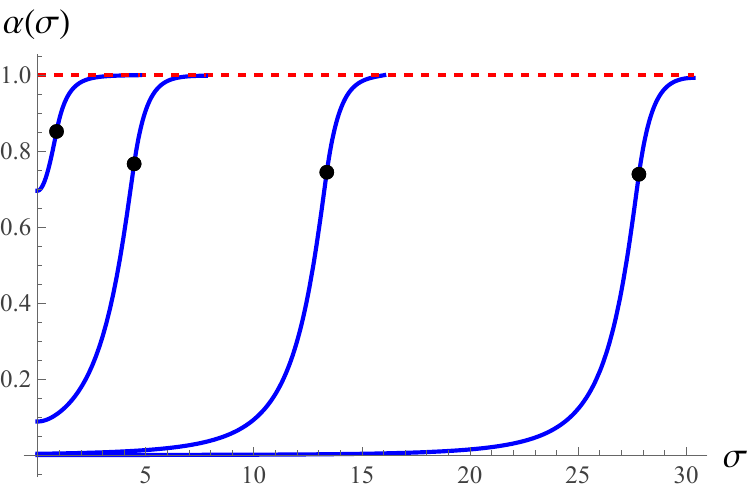}
\caption{On the left, a typical profile of $\alpha(\sigma)$ as a function of $\sigma$ ($\tilde{L}=0.655$). The black dashed line corresponds to $\alpha_i=y_{J,f}^{-1}=0.755$. On the right, the profile of $\alpha(\sigma)$ for various values of $\tilde{L}$ (from left to right: $\tilde{L}=0.695$, $\tilde{L}=0.66$, $\tilde{L}=0.65$, $\tilde{L}=0.647$). Black dots represent the transition points from disconnected to connected phase.}
\label{fig:alpha.profile}
\end{figure}

\begin{figure}[htbp]
\center
\includegraphics[scale=0.6]{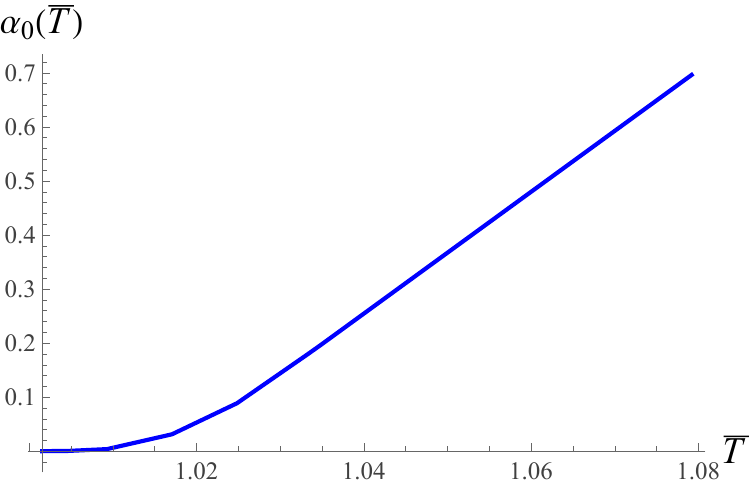}
\includegraphics[scale=0.6]{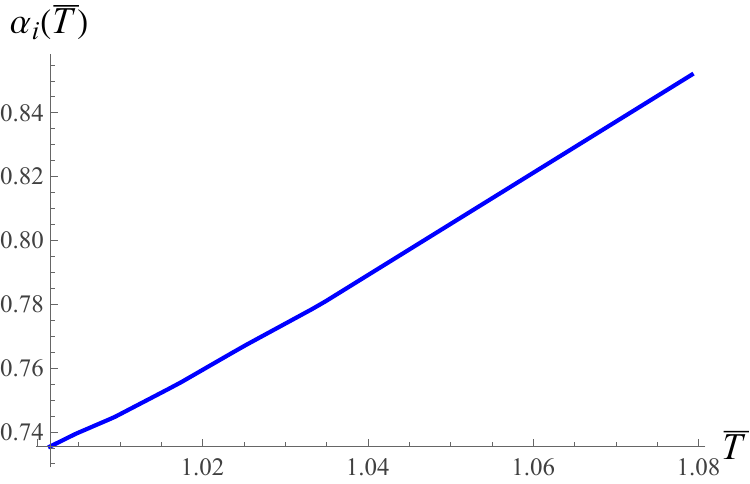}
\caption{Values of $\alpha_0$ and $\alpha_i$ when varying $\bar{T}$.}
\label{fig:alpha0}
\end{figure}

\begin{figure}[htbp]
\center
\includegraphics[scale=0.54]{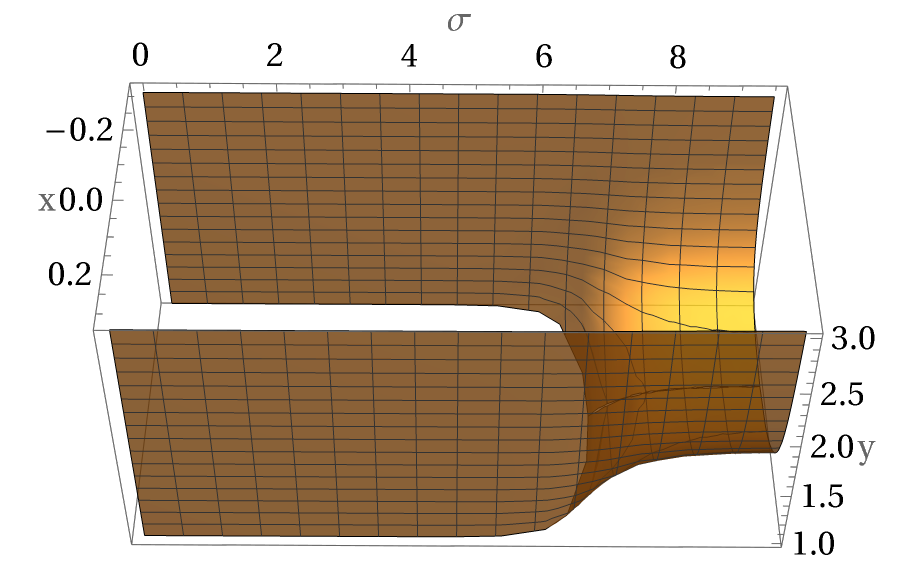}
\includegraphics[scale=0.54]{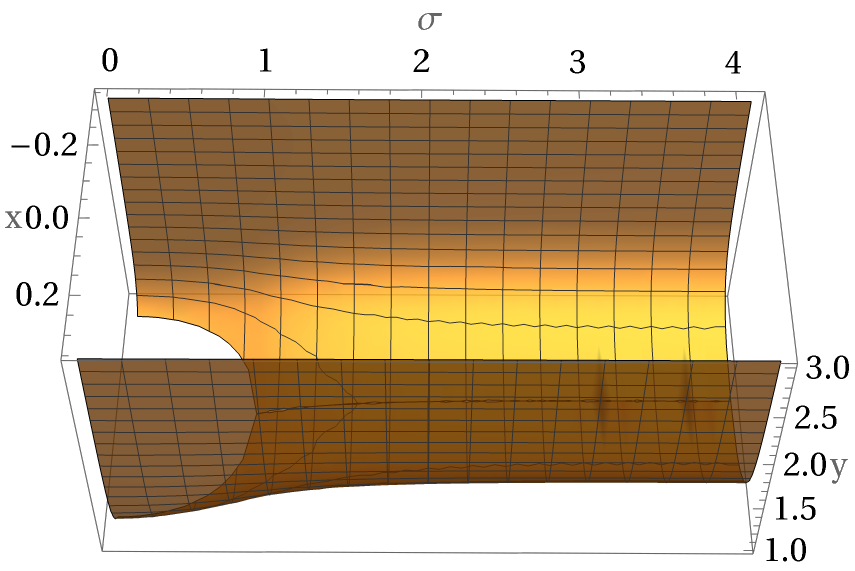}
\caption{The bounce solution for $\tilde{L}=0.655$ is shown on  the left, and that for  $\tilde{L}=0.695$ on the right. The horizontal axis denotes the radial coordinate $\sigma$, the vertical axis is the holographic direction $y$, and the remaining axis corresponds to $x$. 
For smaller $\tilde{L}$, closer to the critical value, most of the interior of the bubble is occupied by an almost vertical configuration. For larger $\tilde{L}$, instead, an almost vertical configuration never appears: the profiles correspond to disconnected configurations that end non-orthogonally on the horizon before becoming connected. }
\label{fig:cartoons}
\end{figure}

We have scanned several values of $\tilde{L}$ over the range \eqref{eq:range} as shown in Figure \ref{fig:alpha.profile}. The shooting parameter $\alpha_0$ is found to be close to zero for values of $\tilde{L}$ near the symmetry-breaking temperature, as illustrated in figure \ref{fig:alpha0}. This behavior is consistent with the expectation that when the temperature is close to the phase transition, the system approaches the thin-wall regime, and the bubble center is very close to the true-vacuum value. For larger values of $\tilde{L}$, however, $\alpha_0$ can deviate significantly from zero. In this case, the bubble's center does not initially coincide with the true vacuum. It is expected that, only after some time from nucleation, it will approach that value. Examples of the brane-antibrane bubble profiles as functions of $\sigma$ can be found in figure \ref{fig:cartoons}.

\begin{figure}[htbp]
\center
\includegraphics[scale=0.6]{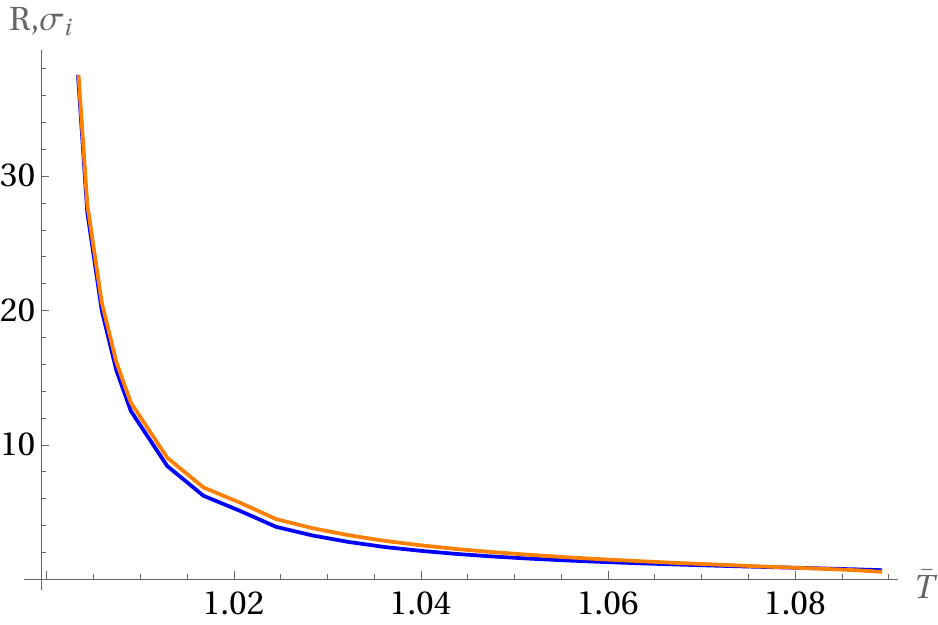}
\caption{Bubble radius $R$ (blue line) and $\sigma_i$ (orange line) as functions of $\bar T$.}
\label{fig.R}
\end{figure}

\begin{figure}[htbp]
\center
\includegraphics[scale=0.7]{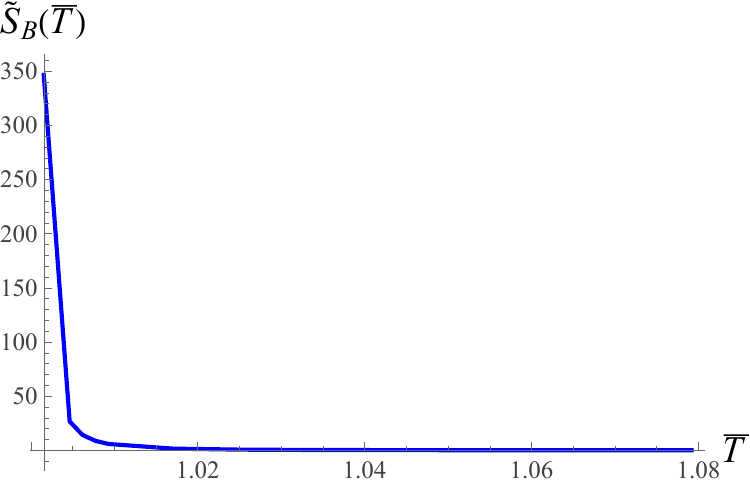}
\caption{The on-shell bounce action $\tilde{S}_{B}[x_B]$ as a function of $\bar T$.}
\label{fig:bounce.action}
\end{figure}

The dimensionless radius $R$ of the bubble can now be defined as the value of $\sigma$ for which $\alpha$ is halfway between its value at the center and its value at the false vacuum
\be
\alpha(R)=\frac{1}{2}\left(\alpha_0+1\right)\,.
\ee
The value of $R$ is very close to $\sigma_i$ as can be seen in figure \ref{fig.R}. 
The Euclidean on-shell bounce action \eqref{eq:bounce} can be computed using the solution for $\alpha(\sigma)$ in \eqref{vari_profile3}  for any given $\tilde{L}$. We plot its value in figure \ref{fig:bounce.action}.

\begin{figure}[htbp]
\center
\includegraphics[scale = 0.64]{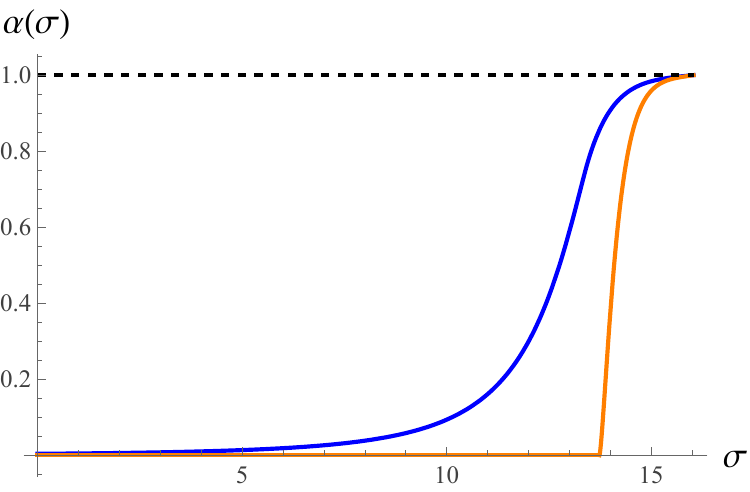}
\includegraphics[scale = 0.64]{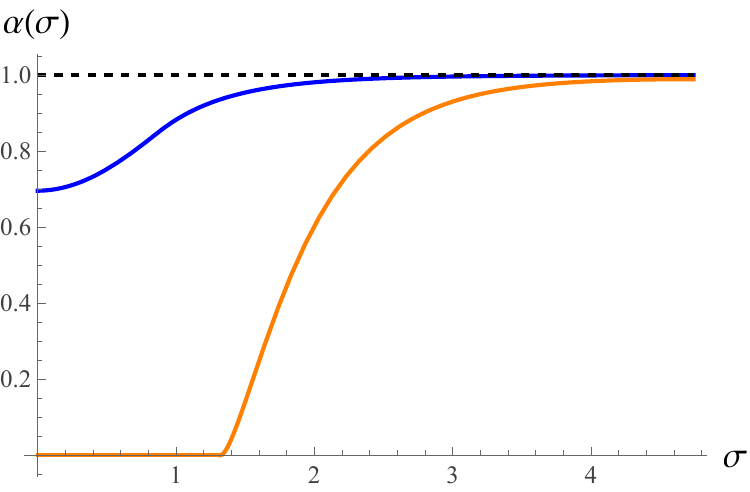}
\caption{Comparison plot between the approximate bounce solution (orange) and the bounce solution in the fully variational case (blue), for $\Tilde{L} = 0.65$ (left) and $\Tilde{L}=0.695$ (right). In order to compare the two bounce solutions, we have shifted the radius coordinate of the approximate one by a factor $\gamma = \frac{\sigma_{c}^{full}}{\sigma_{c}^{approx}}$ where  $\sigma_{c}$ is the finite cutoff  introduced by the numerical analysis. For each value of $\tilde{L}$, the value $\sigma_c$ at which $\alpha(\sigma_c)=1$ differs for the two solutions, and therefore the range of $\sigma$, needs to be adjusted in order to perform a comparison. 
} 
\label{fig: bounceconfr}
\end{figure}

\begin{figure}[htbp]
\center
\includegraphics[scale=0.64]{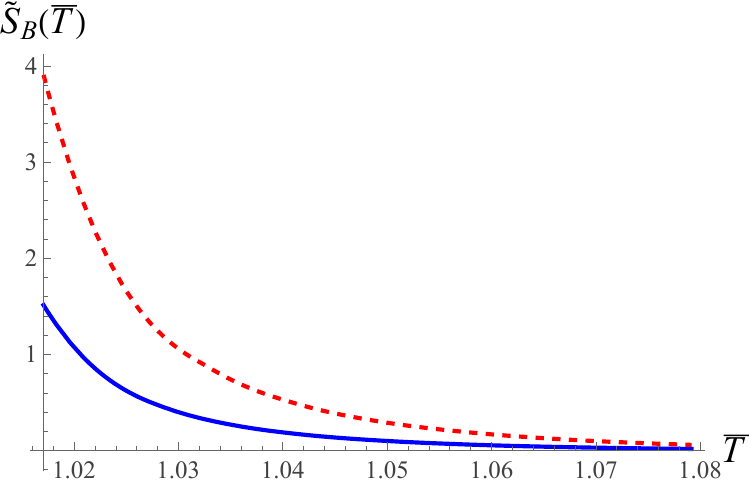}
\includegraphics[scale=0.64]{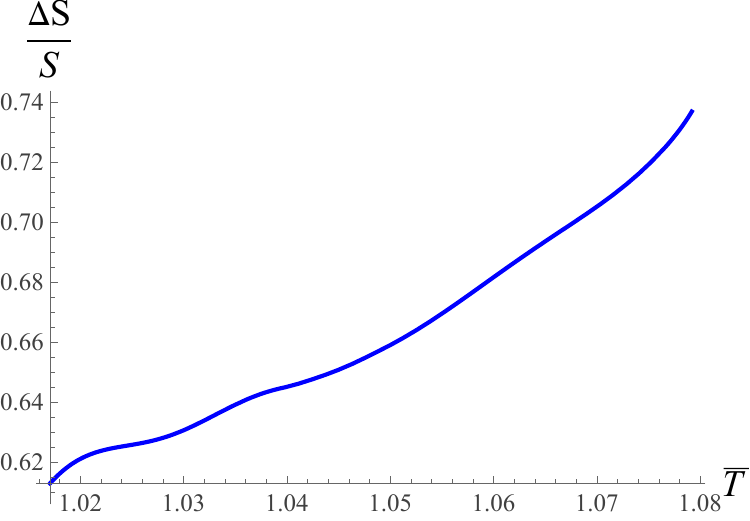}
\caption{The on-shell bounce action for the approximated (dashed) and fully variational (solid) solution on the left and the relative error between the two actions on the right as functions of $\bar T$.}
\label{fig:S.confr}
\end{figure}

To conclude this section, let us compare the fully variational solution with the approximated one found in the previous section. 
Examples of profiles $\alpha(\sigma)$ in the two cases are provided in figure \ref{fig: bounceconfr}.
It is also possible to compare the on-shell bounce actions. The full solution has a smaller action than the approximated solution in the whole range of $\bar{T}$ as can be seen in figure \ref{fig:S.confr}. Moreover, given that the approximated solution captures the full solution better for temperatures close to the critical one, the relative error of the on-shell actions tends to be smaller.
Overall, the full solution provides sizable corrections to the observables calculated with the approximated solution. 

\subsection{Bubble nucleation rate}
\begin{figure}[htb]
\center
\includegraphics[scale=0.7]{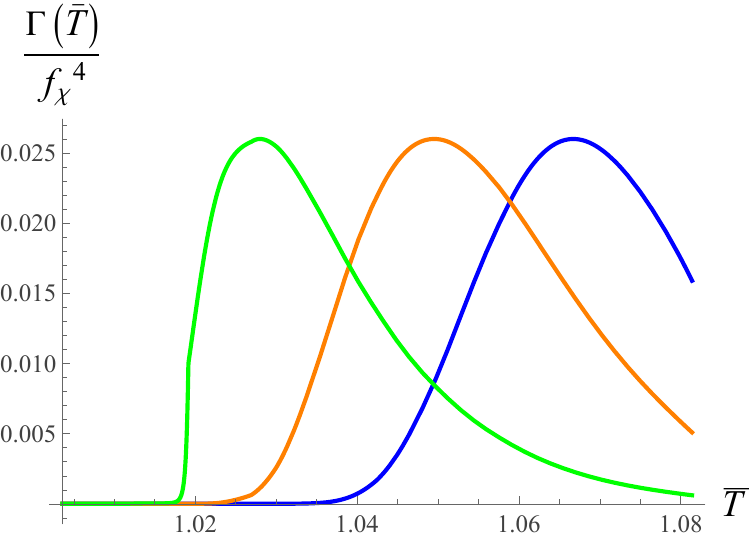}
\caption{The dimensionless bubble nucleation rate as a function of $\Bar{T} $ for $N=3$ and different choices of the parameters $\lambda$, $M_{KK}$ and $L$. Here we have chosen $L$ = $0.2 \frac{\pi}{M_{KK}}$ and the following values of ($M_{KK} (\rm{MeV})$, $\lambda$): (700, 70) (blue line), (600, 50) (orange line) and (500, 30) (green line).}
\label{fig:dbirate}
\end{figure}

For the chiral symmetry restoration transition, the bubble nucleation rate is given by
\be
\label{eq:GamDBI}
\Gamma =f_{\chi}^4 \left( \frac{S_{3,B} }{2\pi T} \right)^{3/2} e^{-S_{3,B} /T}\,,
\ee
where, using \eqref{eq:defiS}, 
\be
\frac{S_{3,B}}{T}\equiv \frac{S_3[x_B]-S_3[x_f]}{T} = \tilde K (\tilde S [x_B]-\tilde S[x_f])\,,
\ee
with $\tilde K$ defined in \eqref{eq:deftildeK}. The order-of-magnitude estimate of the prefactor in \eqref{eq:GamDBI} is due to the fact that now the false vacuum is connected.\footnote{In the supercooled case, instead, where the false vacuum was the disconnected one, the prefactor was estimated as $T^4$ \cite{Bigazzi:2020phm}.} Here
\be
f_{\chi}^2 \approx 0.1534 \frac{\lambda N}{32\pi^3} \frac{1}{M_{KK} L^3}\,,
\ee
is the chiral symmetry breaking scale. Plots of the dimensionless rate $\tilde \Gamma \equiv \Gamma/f_{\chi}^4$ for some benchmark values of the parameters are given in figure \ref{fig:dbirate}.

\subsection{Transition parameters}
The strength parameter of the chiral symmetry restoration transition is defined as in (\ref{defalpha}). In the present case, the energy density ahead of the wall is given by
\be
\rho_+= \rho_{\rm{glue}}+\rho_f\,,
\ee
where 
\be
\rho_{\rm{glue}}=5\frac{2^6 \pi^4}{3^7}\lambda N^2 \frac{T^6}{M_{KK}^2}\,,
\ee
is the energy density of the unflavored plasma in the deconfined phase, while $\rho_f$ is the energy density of the flavor brane sector in the false (i.e.~symmetry broken) vacuum. In the probe approximation the latter is subleading w.r.t.~the former so that we can take $\rho_+\approx \rho_{\rm{glue}}$.

\begin{figure}[htb]
\center
\includegraphics[scale=0.7]{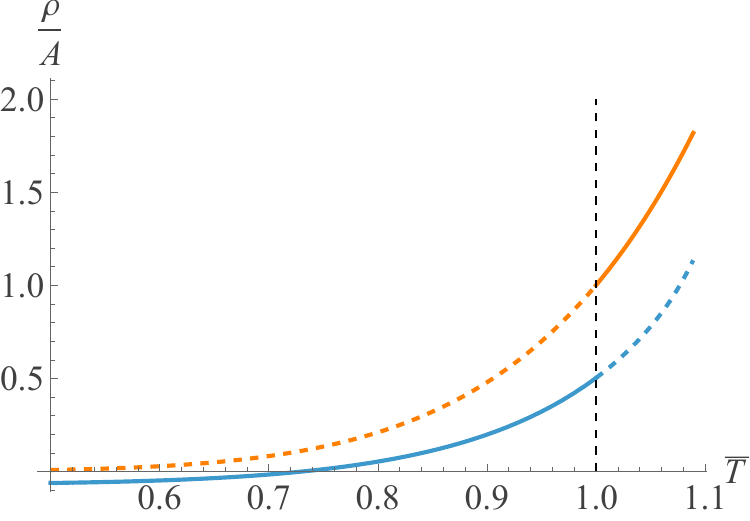}
\caption{The energy density of the flavor sector as a function of the temperature. Solid (resp.~dashed) lines correspond to stable (resp.~metastable) branches. The blue (resp.~orange) line corresponds to the connected (resp.~disconnected) configuration. Here $A=2^6 \pi^2 \lambda^3 N_f N T_c^7(3^7\,7 M_{KK}^3)^{-1}$.}
\label{fig:DBIEnergy}
\end{figure}

In the present setup we have
\be
\theta_f - \theta_t = \frac{\Delta \rho - 3 \Delta p}{4}\,,
\ee
where
\bea
\Delta\rho &=& \rho_f - \rho_t = (1-T\partial_T)\left(\frac{2^3\pi^2}{3^8}\lambda^3 N_f N \frac{T^7}{M_{KK}^3}\Delta\tilde S\right)\,,\nonumber\\
\Delta p &=& p_f - p_t = - \frac{2^3\pi^2}{3^8}\lambda^3 N_f N \frac{T^7}{M_{KK}^3}\Delta\tilde S\,.
\eea 
Here
\be
\label{energyd}
\rho_{t} =\frac{2^6 \pi^2}{7\cdot 3^7} \lambda^3 N_f N \frac{T^7}{M_{KK}^3}\,,\quad p_t = \frac{\rho_t}{6}\,,
\ee
are the energy density and pressure of the disconnected flavor brane configuration. A plot of the energy as a function of the temperature is given in figure \ref{fig:DBIEnergy}.
Hence
\be
\theta_f - \theta_t = - \frac{2\pi^2}{3^8}\lambda^3 N_f N \frac{T^7}{M_{KK}^3}\left(3\Delta\tilde S + T\partial_T \Delta\tilde S\right) <0\,,
\ee
in the range of temperatures we are focusing on. Thus, taking $\rho_+\approx \rho_{\rm{glue}}$, we get
\be
\alpha\approx - \frac{\lambda^2 N_f}{N}\frac{T}{M_{KK}}\frac{\left(3\Delta\tilde S + T\partial_T \Delta\tilde S\right) }{15\pi^2 2^5}\,.
\ee

Again, to specify other transition parameters (and to evaluate $\alpha$ at the nucleation temperature), we need to estimate the typical timescale for the evolution of the whole system. Just as in section \ref{sec:transparcdc}, let us choose to set the latter by $Y$ as given in eq.~\eqref{eq:Yvalue}. Using the same general relations reported in section \ref{sec:transparcdc}, we can thus compute the nucleation temperature and, provided the bubble wall velocity is known, the percolation temperature and the $\beta$ parameter,
useful for obtaining the spectrum of gravitational waves.
Numerical results found using some sample values for the model parameters are shown in table \ref{tab1}.
\begin{table}[h]
\centering
\caption{Values of the transition parameters for the benchmark values of $M_{KK}$ and $\lambda$ used in the graph in figure \ref{fig:dbirate}, with $L=0.2 \pi/M_{KK}$.}
\begin{tabular}{|l|l|c|c|c|c|}
\hline
 & & \textbf{$T_c \; (\rm{MeV})$} & \textbf{$T_n \; (\rm{MeV}) $} & \textbf{$T_p \; (\rm{MeV})$} & \textbf{$\beta/Y$} \\ \hline
\textbf{$M_{KK}=700\ \rm{MeV}$} & \textbf{$\lambda = 70$} & $171.346$ & $173.452$ & $173.511$ & $157.946$\\ \hline
\textbf{$M_{KK}=600\ \rm{MeV}$} & \textbf{$\lambda = 50$} & $146.868$ & $147.911$ & $147.931$ & $156.785$\\ \hline
\textbf{$M_{KK}=500\ \rm{MeV}$} & \textbf{$\lambda = 30$} & $122.39$ & $122.768$ & $122.772$ & $152.983$\\ \hline
\end{tabular}
\label{tab1}
\end{table}
To compute the percolation temperature from equation (\ref{percol}), we have used the bubble wall velocity $v$ found in section \ref{sec:velocity}. 

\section{Bubble wall velocity}
\label{S:vel}

In this section, we consider the asymptotic steady state of the bubble at late times in order to extract its velocity.
This state is reached when the pressure difference between the inside and outside of the bubble is exactly compensated by the friction exerted by the plasma - a ``zero force'' condition.
The discussion will follow the lines of that for the supercooled phase transition in \cite{Bigazzi:2021ucw}.

Since the steady state is a late-time configuration, when the bubble is expected to be large, we consider a planar bubble wall, with $z$ denoting the Minkowski direction of its motion. The steady-state ansatz for the $D8$-brane profile is taken to be
\be
x_4(t,z,u) = x_4 (z-vt, u)\,,
\ee
or, equivalently
\be
z = vt + \xi(u,x_4)\,.
\ee
Also in this case, finding an exact solution to the equations of motion from the DBI action is extremely difficult.
Thus, we will employ a ``rectangular approximation'', where the profile of the $D8$-brane along $x_4$ is taken to be constant, so that $\xi=\xi(u)$.
This is surely a very rough approximation but, as we will see, the results for the friction force and the terminal velocity have a very clear interpretation.
We consider this fact as a signal that the approximation captures the basic properties of the full solution.

\begin{figure}
 	\center
 	\includegraphics[scale=0.45]{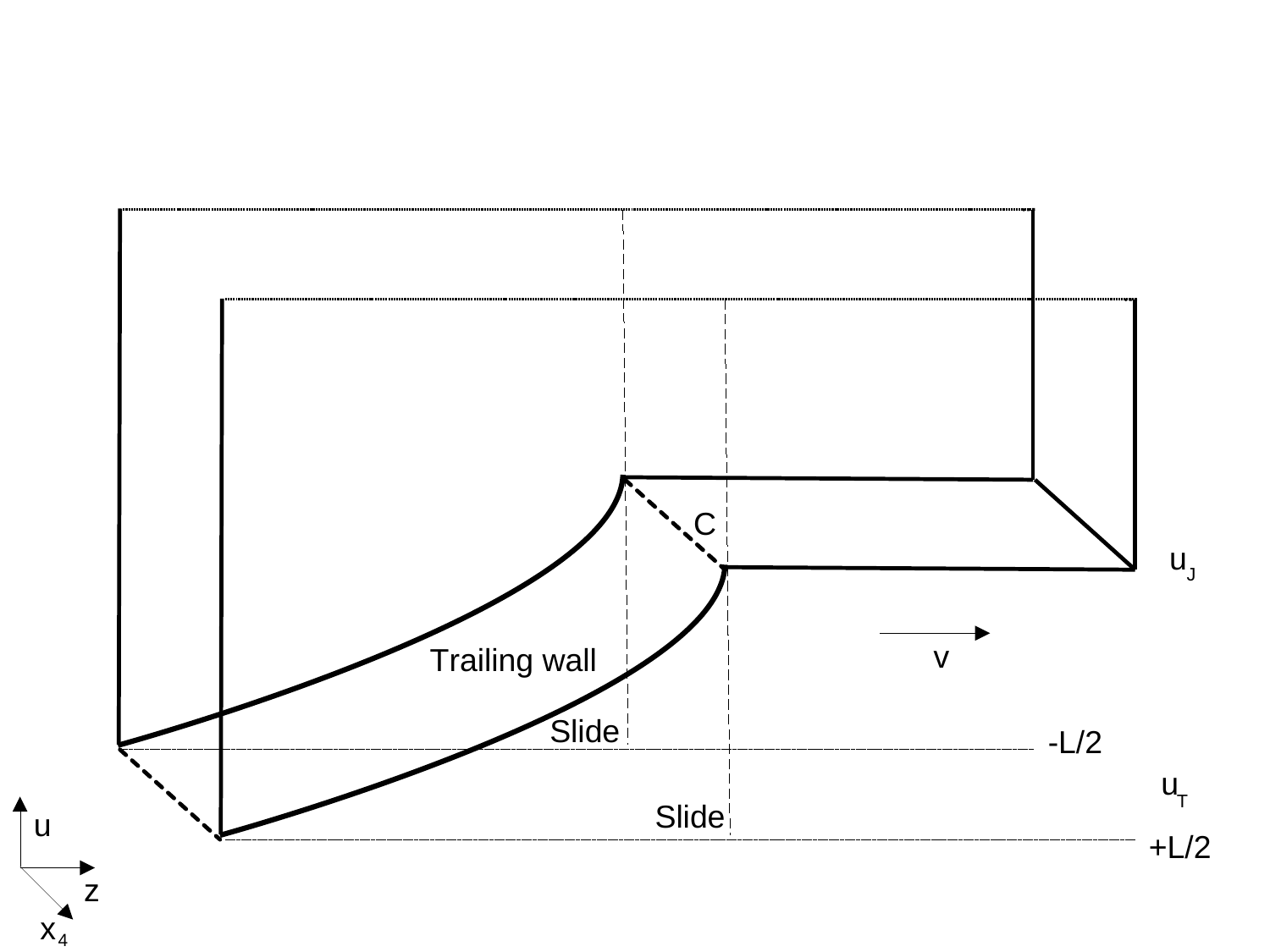}
 	\caption{Cartoon of a possible steady-state bubble in the rectangular approximation. Note that the curve $C$ comprises the horizontal dashed segment at $u_J$ as well as the two dashed vertical segments from $u_J$ to infinity.}
 	\label{fig:rectangular}
 \end{figure}

The steady state configuration we  consider interpolates between a disconnected embedding at $z \rightarrow -\infty$ and a connected one at $z \rightarrow +\infty$, and it  looks as in figure \ref{fig:rectangular} in the rectangular approximation.
This displays a non-smooth connection of the ``trailing wall'' - the part of the brane leaning from the connected configuration towards the horizon - to the rest of the configuration, supported by the equilibrium of the various forces.
The cartoon in figure \ref{fig:rectangular}, in any case,  schematically captures a salient feature expected from the steady state configuration, i.e.~the fact that the metastable phase is sucked in by the expanding bubble. This is a generic feature of inverse phase transitions, in which energy is drawn from the metastable plasma into the bubble interface, thereby favoring plasma motion toward the center of the bubble as in \cite{Barni:2025gnm}.

\subsection{Warm-up: the trailing wall}
\label{sec:trailing}

Let us first consider just the ``trailing wall'', i.e.~the part of the brane which trails beyond the wall, connecting to the false-vacuum configuration and asymptotically touching the horizon at $z \rightarrow -\infty$.
It is the most important piece of the profile because it is responsible for exchanging momentum with the horizon, causing friction similarly to a trailing string \cite{Herzog:2006gh,Gubser:2006bz}.
As in figure \ref{fig:rectangular}, here we envisaged the simplest possibility: that the trailing wall has a monotonic profile. 

The equation of motion for the profile of the trailing wall of the steady state, which neglects the dependence on $x_4$ and can be parameterized as
\be 
z_w= vt +\xi(u)\,,
\ee
can be derived from the DBI action
\be
\label{eq:Sw}
S_w = -\frac{k}{L} \int dt\, du\, dx_4\, u^4 \left(\frac{R}{u} \right)^{3/2} \sqrt{1+f_T(u)\left(\frac{u}{R} \right)^{3}(\partial_u \xi)^2-v^2/f_T(u)}\,,
\ee
with
\be\label{kappa}
k = \frac{T_8}{g_s} A L V(S^4)\,, \qquad A = \int dx_1 dx_2\,,
\ee
and it is given by
\be\label{eqrectangular}
\xi' = \pi^u_\xi \left( \frac{R}{u}\right)^{3/2} \frac{1}{\sqrt{f_T(u)}} \sqrt{\frac{1-\frac{v^2}{f_T(u)}}{k^2 f_T(u) u^8-(\pi^u_\xi)^2}}\,,
\ee
where the conjugate momentum $\pi^u_\xi$ is the world-volume current of spacetime energy-momentum
carried by the brane.
If we just focus on the trailing wall, neglecting its interaction with the rest of the configuration,  $\pi^u_\xi$ is a constant ($\partial_u \pi^u_\xi =0$) in the rectangular approximation.
Note that the derivative of $\xi$ goes to infinity at the horizon (``trailing'' profile).

The right-hand side of equation (\ref{eqrectangular}) is real for every value of $u$ if both the numerator and the denominator under the square root change sign at 
\be
u_c(v) = \frac{u_T}{(1-v^2)^{1/3}}\,.
\label{uc}
\ee
Correspondingly
\be
\label{eq:horizonc}
\pi_{\xi}^u =  k u_T^4 \frac{v}{(1-v^2)^{4/3}}\,.
\ee
Since $u_T\sim T^2$, the relation (\ref{uc}) can also be read as a definition of the boosted temperature
\be
T_{boost}= \frac{T}{(1-v^2)^{1/6}}\,.
\ee

Precisely as in \cite{Bigazzi:2021ucw}, the drag force is the momentum flow $dp_z /dt$ going down the brane and transferred to the
horizon, 
\be
F_d=\frac{dp_z}{dt}=\pi_{\xi}^u\,.
\ee
Accordingly, from (\ref{eq:horizonc}) and (\ref{kappa}) the drag force per unit area is found to be
\be\label{effesua}
\frac{F_d}{A}= \frac{2^5}{3^9} \pi^3 \lambda^3 N N_f  (L T) \frac{T^7}{M_{KK}^3}\frac{v}{(1-v^2)^{4/3}}\,.
\ee
Taking into account the expression (\ref{energyd}) for the energy density $\rho_t$ of the disconnected configuration (which is the true vacuum), the drag force can also be expressed as
\be
\frac{F_d}{A} = \frac{\pi}{3} L\, T_{boost}\, w_{t}(T_{boost})\, v \equiv C_d \,\frac{T_{boost}}{T^\chi_c}\, w_{t}(T_{boost})\, v\,,
\label{fdragexpr}
\ee
where $C_d\approx 0.16$ is a model-dependent drag coefficient \cite{Bigazzi:2021ucw}, $T^\chi_c=0.1538 L^{-1}$ is the critical temperature for the phase transition and $w_{t}(T_{boost})$ is the enthalpy density ($w= \rho+ p$) of the true vacuum at the boosted temperature $T_{boost}$. 
Notice that it is the enthalpy of the \emph{true vacuum} which provides the friction. Instead, in the supercooled case, the drag force turned out to be proportional to the enthalpy of the \emph{false vacuum}  \cite{Bigazzi:2021ucw}. 

\subsection{The complete steady state configuration} 

In this section, we consider the full steady state configuration, comprising the trailing wall and the rest of the brane profile, to be determined dynamically taking the boundary conditions into account.
In the rectangular approximation, the whole configuration, including the trailing wall, is rigid along the $x_4$ direction. Moreover, the fixed $z$ slices of the connected part (the false vacuum outside the bubble) is composed of two vertical lines at $x_4 = \pm L/2$ and a horizontal line at $u=u_J$.
The steady state is completed by vertical pieces connecting to the trailing wall inside the bubble, as in figure \ref{fig:rectangular}.

We analyze the equations of motion for the wall, exploiting the results of \cite{Bigazzi:2021ucw}. Compared to the supercooled case analyzed in  that paper, one can see that the vertical parts of the brane inside the bubble miss the two vertical slides shown in yellow in figure 3 of \cite{Bigazzi:2021ucw}, which are the ``Slide'' pieces below the ``Trailing wall'' in figure \ref{fig:rectangular}. We thus subtract the corresponding contributions from the total action, which we write as
\be
\label{eq:action.full}
S = S_{conn} + S_{disc} + S_w - S_{sl}  \ ,
\ee
where
\begin{subequations}
\ba
S_{conn} &=&  -\frac{k}{L} \int dt \int dx_4 \int_{u_T}^{+\infty} du\, \delta(u-u_{ss}) \int_{z_w(C)}^{\infty}dz \,{\cal L}_{c}\ , \\
S_{disc} &=&  -\frac{2 k}{L} \int dt \int dx_4 \, \delta(x_4-L/2) \int_{u_T}^{+\infty} du  \int_{-\infty}^{z_w(C)} dz \, \,{\cal L}_{d}\ ,\\
S_{sl}& =& - \frac{2k}{L}\int dt \int dx_4\, \delta(x_4-L/2) \int_{u_T}^{+ \infty} du  \int_{z_w(u,L/2)}^{z_w(C)} dz{\cal L}_d  \ , \\
\label{Sslides2b}
S_{w} &=& - \frac{k}{L}\int dt \int dx_4 \int_{u_T}^{+\infty}  du \, \Theta (u_{ss}-u)  {\cal L}_w(\partial z_w) \ .
\ea
\end{subequations}
The notation in these formulas is the following:  $z_w$ is the profile of the bubble wall ($C$ being the curve where it attaches to the rest of the configuration), $S_{conn,disc}$ (${\cal L}_{c,d}$) are the actions (Lagrangians) of the connected and disconnected configurations that can be obtained from \eqref{SDBI1}. Notice that $S_{conn,disc}$ depend on $z_{w}$ only through the limits of integration in $z$ while ${\cal L}_{c,d}$ do not depend on it. $S_w$ (${\cal L}_w$) is the action (Lagrangian) of the wall given in \eqref{eq:Sw}, and $S_{sl}$ is the action of  the vertical slides, which in the present case are subtracted from the complete configuration. 
The domain of the trailing profile $z=z(u,x_4)$ is the region $D_C$ spanned by $x_4 \in [-L/2,L/2]$ and $u \in [u_T,C]$, where $C$ is the curve $u=u_{ss}$, which describes the connected solution.

Taking the variation of the action \eqref{eq:action.full} with respect to $z_w (u)$, we get
\ba
\label{variationtotalactionnonrectangular}
\d S &=& \frac{1}{L} \int dt \int dx_4 \int_{u_T}^{+\infty}  du\,  \Theta (u_{ss}-u) (\p_u \pi_\xi ^u) \, \d z_w (u) \nb \\
&-& \frac{1}{L} \int dt \int dx_4 \int_{u_T}^{+\infty} du\, \delta(u-u_{ss}) \pr{-k \mc{L}_c + \pi_\xi ^u} \d z_w (C) \nb \\
&+&\frac{1}{L} \int d t \int dx_4 \pi_\xi ^u |_{u=u_T} \d z_w (u_T) \nb \\
&-& \frac{1}{L} \int d t \int_{u_T} ^{+\infty} du \pq{ k \mc{L} _d}_{x_4=L/2} \d z_w (u,L/2) \nb \\
&+& \frac{1}{L} \int d t \int_{u_T} ^{+\infty} du \pq{ - k  \mc{L} _d}_{x_4=-L/2} \d z_w (u,-L/2)  \ ,
\ea
where $\pi_{\xi}^u=\frac{\partial }{\partial \xi'(u)}k{\cal L}_w$.

The third line of (\ref{variationtotalactionnonrectangular}) is vanishing upon imposing Dirichlet conditions at the horizon, while
from the last two lines we see that there is a source term in the equation of motion. 
In principle, this is a source localized at $x_4 = \pm L/2$. The rectangular approximation consists in taking this source as independent of $x_4$ since the tension of the brane along $x_4$ becomes infinite. As a result, the 
Euler-Lagrange equation on $D_C$ is
\be
\partial_u\pi_{\xi}^u =   2 k {\cal L}_d \,.
\label{eqpid0}
\ee
The equation of motion is supplemented with the boundary condition coming from the
second line of (\ref{variationtotalactionnonrectangular}), 
\be
\int dx_4 \int_{u_T}^{+\infty} du\, \delta(u-u_{ss}) \pi_\xi ^u (x_4,u) =  k \int dx_4 \int_{u_T}^{+\infty} du\, \delta(u-u_{ss}) \mc{L}_c \ ,
\label{bdruj}
\ee
which in the rectangular approximation reduces to
\be
\label{bconditionatujrectangular}
\pi_\xi ^u (u_J) =  k u_J^4 \sqrt{f_T(u_J)}\equiv  k \mc{L}_{ch}(u_J)\,.
\ee
Here one can recognize a term proportional to the Lagrangian density of the horizontal part of the  connected configuration (${\cal L}_{ch}$) in the rectangular approximation, trivially extended along $x_4$ at $u=u_J$. 

Integrating the Euler-Lagrange equation (\ref{eqpid0}) over the whole $D_C$, we find
\be
\label{g1nonrectangular}
\frac{1}{L}\int dx_4\, \pi_{\xi}^{u}(u_T, x_4) = -\frac{2k}{L}\int_{u_T}^{+\infty}du{\cal L}_d + \frac{k}{L} \int dx_4 \int_{u_T}^{+\infty} du\, \delta(u-u_{ss}) \mc{L}_c \ .
\ee
The right-hand side of eq.~(\ref{g1nonrectangular}) is proportional to the difference between the static connected and disconnected on-shell DBI Euclidean actions. In particular
\be
T\Delta S =T(S_c-S_d) = - \frac{2k}{A L} \int d^3 x \int_{u_T}^{+\infty}du{\cal L}_d + \frac{k}{A L} \int d^3 x \int dx_4 \int_{u_T}^{+\infty} du\, \delta(u-u_{ss}) \mc{L}_c \ ,
\ee
where $A=\int d^2x$. 
Using the holographic relation between the above expression and the free energy in the dual QFT, we get
\be
T \Delta S = \Delta {\cal F} = \int d^3x \Delta f = - \int d^3 x (p_c-p_d) \equiv  \int d^3 x (p_t-p_f) \equiv \int d^3 x \Delta p\,,
\ee
where $f$ is the free energy density and $\Delta p$ is the pressure difference between the true and false vacua.

Thus, equation (\ref{g1nonrectangular}) reads
\be
F\equiv\overline{\pi_{\xi}^u(u_T)} \equiv \frac{1}{L}\int dx_4\, \pi_{\xi}^{u}(u_T, x_4)  =  A\, \Delta p\,.
\label{expenonrect}
\ee
The left-hand side of equation (\ref{expenonrect}), i.e.~the (average along $x_4$) of the momentum flow towards the horizon, is interpreted as the \emph{total friction force} that the plasma exerts on the wall. Thus, equation (\ref{expenonrect}) relates the friction force to the pressure difference, and therefore gives the zero-force condition for the steady state.
As we will see in the next section, the total friction force can be written as the sum of the drag force $F_d$ computed in section \ref{sec:trailing} and another contribution. 
In \cite{Bigazzi:2021ucw} it was shown that the zero-force condition is valid beyond the rectangular approximation employed here.

\subsection{The velocity}\label{sec:velocity}
Once the zero-force condition equation (\ref{expenonrect}) is derived, one needs to evaluate its left-hand side, namely the total friction force, since its right-hand side is known. 
To this aim we note that (\ref{eqpid0}) can be written explicitly as
\be
\partial_u\pi_{\xi}^u =  2\frac{k}{L} R^{3/2} u^{5/2}\,.
\label{eqpid}
\ee
Integrating (\ref{eqpid}), we get
\be
\pi_{\xi}^u = \frac47 \frac{k}{L} R^{3/2} ( u^{7/2} - u_*^{7/2})\ .
\label{pinte}
\ee
Using the boundary condition (\ref{bconditionatujrectangular}), from (\ref{pinte}) we get 
\be \label{qui0}
u_*^{7/2} = u_J^{7/2} - \frac{7L}{4 R^{3/2}}{\cal L}_{c h}(u_J) \,,
\ee
which in turn can be rewritten as
\be 
\label{ustar}
u_*^{7/2} = u_T^{7/2} - \frac{7L}{4k R^{3/2}} \, A\Delta p \,,
\ee
where again, $\Delta p = p_d-p_c \equiv p_t- p_f$, which in the rectangular approximation reads 
\be \Delta p|_{rect} = -\frac47 \frac{k R^{3/2}}{A L}\left( u_J^{7/2} - u_T^{7/2}\right) + \frac{k}{A} \sqrt{f_T(u_J)} u_J^4\,.
\label{rectangulardeltap}
\ee
Notice that this expression has to be handled with care. In particular, if we want it to mimic, at least qualitatively, the actual expression for $\Delta p$ obtained from the smooth configuration (see figure \ref{fig:deltaStilde} and eq. \eqref{deltaptrue} below), we have to artificially tune $L$ in such a way that $\Delta p|_{rect}$ displays two zeros as a function of $u_J$: the trivial one at $u_J=u_T$ and the ``critical" one at the value of $u_J/u_T$ corresponding to the critical temperature $T_c$ below which $\Delta p$ becomes negative.\footnote{For instance, setting $\tilde L=0.415$, the function in \eqref{rectangulardeltap} starts from zero at $u_J=u_T$, then reaches a maximum at $u_{J,m}/u_T\sim 1.06$ and decreases, crossing zero at $u_J/u_T\sim 1.35$. The unstable branch corresponds to the values $u_J\in [u_T, u_{J,m}]$.} 

\begin{figure}
	\center
	\includegraphics[scale=0.7]{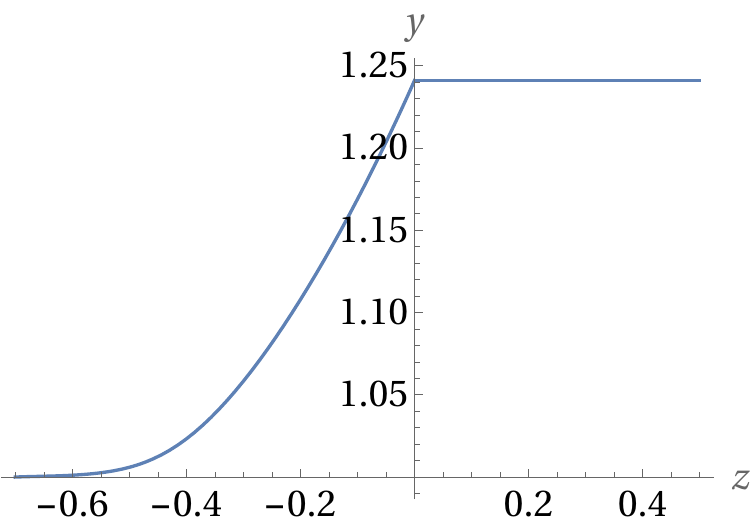}
	\caption{Plot of the trailing wall along the holographic radial direction $y$ for $\bar T=1.05$, corresponding to the value of the velocity $v=0.111$.}
	\label{fig:profile}
\end{figure}

 We are now in a position to derive the formula for the steady-state velocity. 
 Using expression (\ref{pinte}), formula (\ref{eqrectangular}) for
 the trailing wall profile is now given by
\be 
\xi' =  \frac47 \frac{k R^{3/2}}{L}\left( u^{7/2} - u_*^{7/2}\right)\left(\frac{u}{R}\right)^{-3/2}f_T(u)^{-1/2}\sqrt{\frac{1-f_T(u)^{-1}v^2}{k^2 f_T(u) u^8 - \frac{16}{49} \frac{k^2 R^{3}}{L^2}\left( u^{7/2} - u_*^{7/2} \right)^2}}\,.
\label{xiprimenew}
\ee 
In figure \ref{fig:profile} we plot a sample profile of the trailing wall, solution of equation (\ref{xiprimenew}), corresponding to a fixed-$x_4$ section of the cartoon in figure \ref{fig:rectangular} at fixed $t$.

At the radial position $u_c$ where the numerator and the denominator in the square root of (\ref{xiprimenew}) are zero, one has\footnote{Notice that, due to the boundary condition \eqref{bconditionatujrectangular}, in the rectangular approximation the denominator would also vanish in $u_J$. However, taking care of the previous observations concerning $\Delta p|_{rect}$, for velocities solving the zero-force condition \eqref{effesuahere}, $u_J$ and $u_c$ would consistently remain different, with $u_J>u_c$.}
\be
k f_T^{1/2}(u_c) u_c^4 = \frac47 \frac{k R^{3/2}}{L} (u_c^{7/2} - u_*^{7/2}) \,.
\ee
Taking equation (\ref{expenonrect}) into account and making use of (\ref{ustar}), we get the zero-force condition
\be\label{effesuahere}
\frac{F}{A}=\Delta p= \frac{2^5}{3^9} \pi^3 \lambda^3 N N_f  (L T) \frac{T^7}{M_{KK}^3} \left[ \frac{v}{(1-v^2)^{4/3}}  +\frac47 \frac{3}{4\pi L T}\left( 1- \frac{1}{(1-v^2)^{7/6}} \right)\right]\,,
\ee
which is the implicit equation for the steady state velocity, providing, at the same time, the expression for the total friction force per unit area $F/A$. 
The first term on the right hand side is the drag force per unit area $F_d/A$ (see eq.~\eqref{effesua}), while the second one is a correction to this force coming from the loss (or gain) of pressure behind the trailing wall.

Notice that
\be
\Delta p = p_t - p_f \equiv p_{d}- p_{c} = \frac{2^3}{3^8} \pi^2 \lambda^3 N N_f \frac{T^7}{M_{KK}^3}\Delta\tilde S\,,
\label{deltaptrue}
\ee
where $\Delta\tilde S$ is proportional to the difference between the on-shell DBI action of the connected vacuum and that on the disconnected one.\footnote{Notice that since the results in this subsection have been obtained within the rectangular approximation, one should in principle use for $\Delta p$ its (suitably tuned) expression \eqref{rectangulardeltap}. Here, following the same arguments as in \cite{Bigazzi:2021ucw}, we are extrapolating our results to the actual smooth configuration.} As a result we can rewrite equation (\ref{effesuahere}) as
\be\label{eqforv2}
\Delta \tilde S = \tilde L \left[ \frac{v}{(1-v^2)^{4/3}}  +\frac{4}{7\tilde L}\left( 1- \frac{1}{(1-v^2)^{7/6}} \right)\right]\,,
\ee
which can be numerically solved giving $v$ as a function of
\be
\bar T=\frac{T}{T^{\chi}_c} = \frac{\tilde L}{\tilde L_c} \approx 1.5518\, \tilde L\,.
\ee
Using 
\be
p_t \equiv p_d = \frac{2^5}{3^8\,7} \pi^2 \lambda^3 N N_f \frac{T^7}{M_{KK}^3}\,,
\ee
and \eqref{effesua} in \eqref{effesuahere}, we get
\be
\frac{F}{A}=\frac{F_d}{A}+p_t(T)-p_{t}(T_{boost})\,,
\ee
which, using the expression $F/A=\Delta p=p_t(T)-p_f(T)$, can also be rewritten as
\be\label{neat}
\frac{F_{d}}{A} = p_t (T_{boost}) - p_f(T)\,.
\ee
Equivalently, using eq.~(\ref{fdragexpr}), the implicit equation determining the velocity can be rewritten as
\be
\boxed{
v = C_d^{-1}\frac{T_c}{T_{boost}}\frac{p_t(T_{boost})-p_f(T)}{w_t(T_{boost})}\,.}
\label{equv}
\ee
Inheriting experience from the supercooled case \cite{Bigazzi:2021ucw}, the above equation is expected to hold, in form, in generic WSS-like $Dp-Dq-\bar{D}q$ setups, with $T_{boost} = T (1-v^2)^{\frac{p-5}{2(7-p)}}$. 

In the small velocity limit, the total friction force is given by
\be
\frac{F}{A}\approx \frac{2^5 \pi^2 \lambda^3 N N_f T^7}{3^9 M_{KK}^3}\left (\pi L T v - \frac{1}{2}v^2\right)\,.
\label{F/Asmallv}
\ee
The ${\cal O}(v)$ drag force contribution in the above expression, which to linear order would give (see also (\ref{equv}))
\be \label{wt}
v \sim \frac{\Delta p}{w_t}\,,
\ee
is complemented by a negative ${\cal O}(v^2)$ 
term which does not explicitely depend on the details of the transition and is proportional to the enthalpy density of the \emph{true vacuum}. 
Notice that this is the opposite of the ``snowplow" effect noticed in the supercooled (sc) case, where 
\be
v_{sc} = C_d^{-1}\frac{T_c}{T_{boost}}\frac{p_t(T)-p_f(T_{boost})}{w_f(T_{boost})}\sim \frac{\Delta p}{w_f} +{\cal O}(v_{sc}^2)\,,
\label{equvsc}
\ee
and the friction depends on the enthalpy density of the \emph{false vacuum} \cite{Bigazzi:2021ucw}. Something similar has been noticed in \cite{Barni:2024lkj,Barni:2025gnm} in the hydrodynamic approximation: the metastable plasma is sucked by the expanding bubble wall during superheated phase transitions, so that the dynamics of the latter acts as a false vacuum cleaner.

The behaviors (\ref{neat}), (\ref{wt}) are particularly neat results coming from this set-up; we would be surprised if they would be drastically spoiled when going beyond the rectangular approximation.

\begin{figure}
	\center
	\includegraphics[scale=0.48]{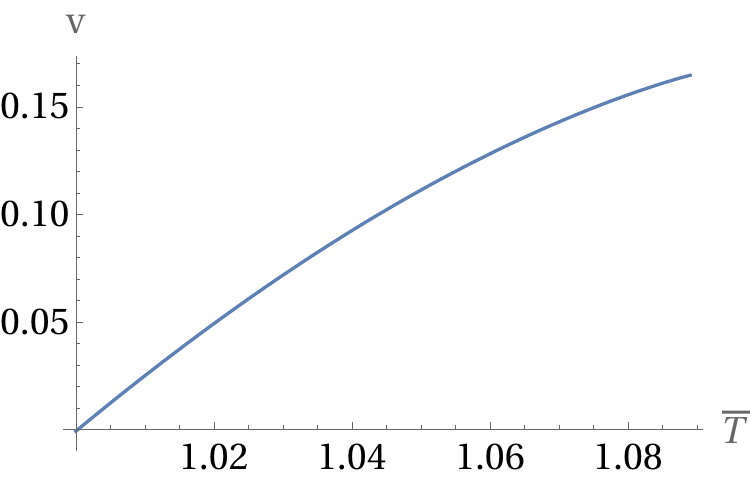}\quad 
    \includegraphics[scale=0.55]{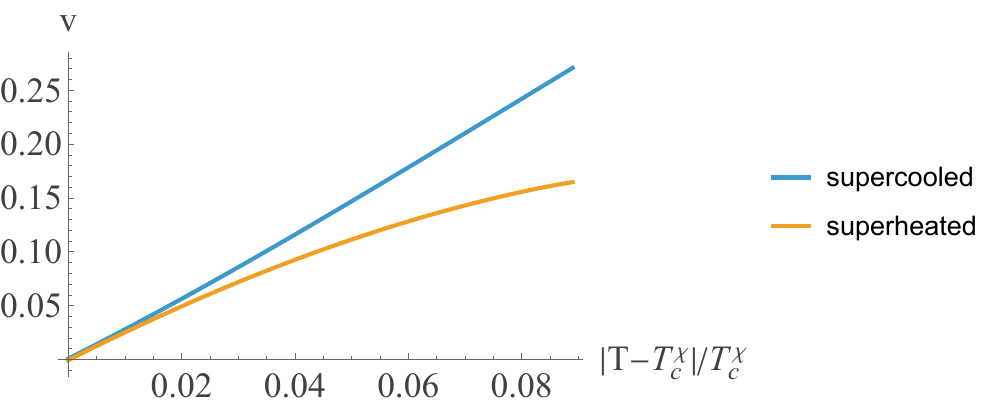}
		\caption{Plot of the terminal velocity of the bubble as a function of $\bar T=T/T^{\chi}_c$ (left), and as a function of the amount of superheating (right, orange curve) or supercooling (right, blue curve).}
	\label{fig:velocity}
\end{figure}

We plot the velocity, obtained by solving equation (\ref{eqforv2}), as a function of $\bar T = T/T_c^\chi$ in figure \ref{fig:velocity} (left).
The connected metastable configuration exists up to a maximum value $\bar T \sim 1.089$.
As a consequence, the velocity has a (rather small) maximal value $v_{max} \sim 0.164$, so our configuration is always subsonic. 
Note, however, that this value is considerably larger than the maximum velocity found in \cite{Bea:2024bls}, which is the only other superheated bubble velocity calculated from first principles in a bottom-up, finite-density holographic model, $v_{max} \sim 0.032$.
This difference would translate into an order of magnitude difference (at least) in the amplitude of the gravitational-wave signal generated by the transition.

In figure \ref{fig:velocity} (right) we compare the velocity of the superheated bubble calculated here with that of the supercooled bubble calculated in \cite{Bigazzi:2021ucw}, for the same amount of superheating/supercooling, i.e.~as functions of $|T-T^\chi_c|/T^\chi_c$.
The superheated bubble is always slower than the supercooled one.
The same behavior was found in \cite{Bea:2024bls}.

\subsection{Alternative configuration} 

\begin{figure}
	\center
	\includegraphics[scale=0.45]{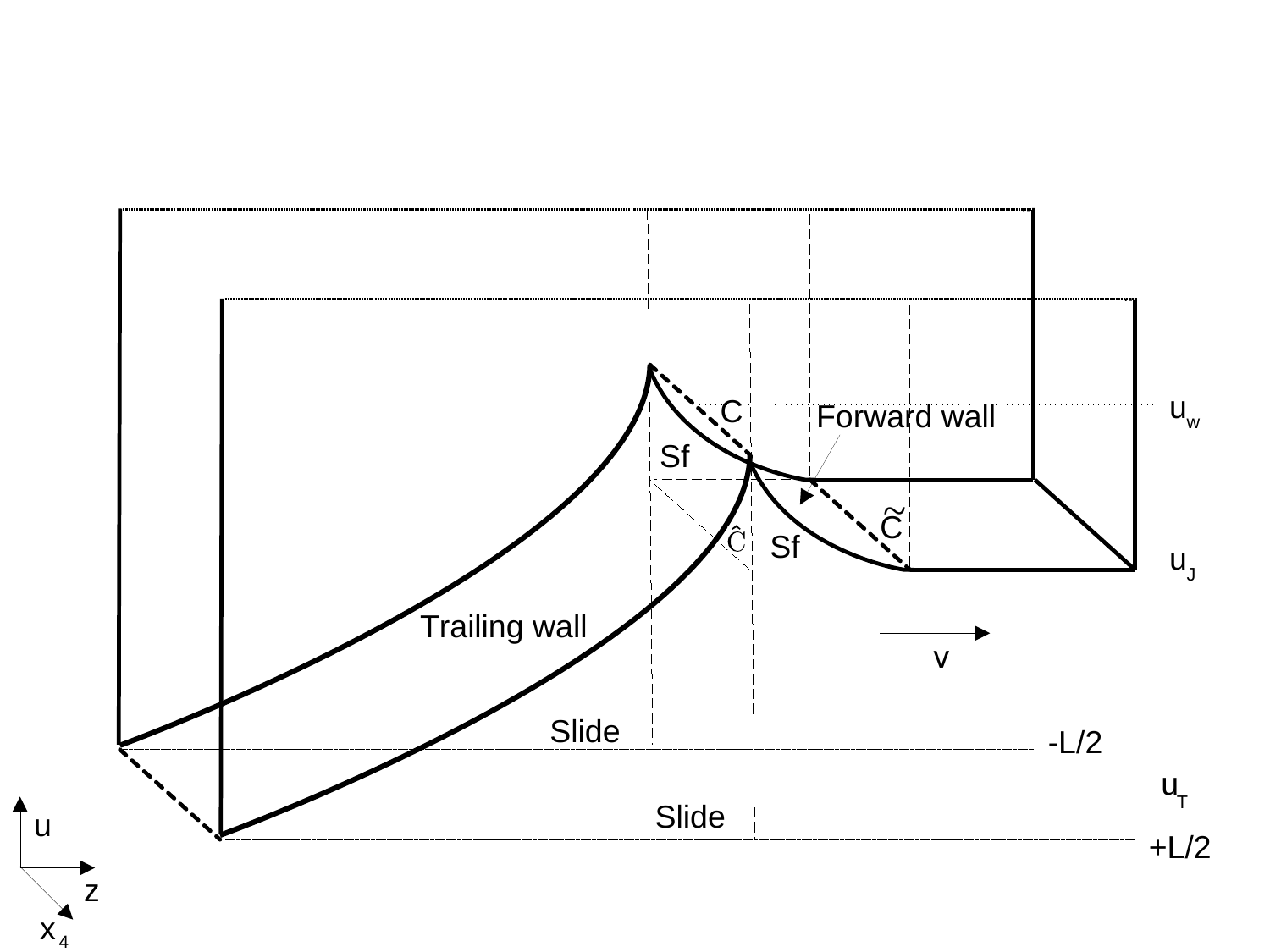}
	\caption{Cartoon of an alternative configuration of the steady-state bubble in the rectangular approximation.}
	\label{fig:fluid}
\end{figure}

In the previous sections we have studied a reasonably motivated ansatz for the steady state.
The considered configuration has a monotonic trailing wall.
One might wander if there exist other simple configurations which are non-monotonic, and what could be the difference in the result for the velocity.
To tackle this question, in this section we consider the alternative configuration in figure \ref{fig:fluid}.
As we will see, the resulting velocity is the same as the one calculated above.

In the new ansatz of figure \ref{fig:fluid}, 
the trailing wall does not join directly to the false vacuum configuration: there is a region in between, which we are going to call ``forward leaning wall'' or ``forward wall'' for brevity, and which moves rigidly with the same velocity of the wall.
So we employ the parametrization
\be 
z = vt + \xi(u)\,,
\ee 
for the trailing wall part ($\xi(u)$ denotes the profile of the trailing wall), and
\be 
z = vt + \xi_{fl}(u)\,,
\ee
for the ``forward'' part ($\xi_{fl}(u)$ denotes the profile of the forward leaning wall).

The procedure to determine the velocity follows the same steps as above, so we will mainly discuss the novel points. 
The action is now
\be
S = S_{conn} + S_{disc} + S_w - S_{sl} +S_{fl} - S_{sf} - S_h \ ,
\ee
where
\begin{subequations}
	\ba
	S_{conn} &=&  -\frac{k}{L} \int dt \int dx_4 \int_{u_T}^{+\infty} du\, \delta(u-u_{ss}) \int_{z(\hat C)}^{\infty}dz \,{\cal L}_{c}\ , \\
	S_{disc} &=&  -\frac{2 k}{L} \int dt \int dx_4 \, \delta(x_4-L/2) \int_{u_T}^{+\infty} du  \int_{-\infty}^{z_w(C)} dz \, \,{\cal L}_{d}\ ,\\
	S_{sl}& =& - \frac{2k}{L}\int dt \int dx_4\, \delta(x_4-L/2) \int_{u_T}^{+ \infty} du  \int_{z_w(u,L/2)}^{z_w(C)} dz{\cal L}_d  \ , \\
	S_{w} &=& - \frac{k}{L}\int dt \int dx_4 \int_{u_T}^{+\infty}  du \, \Theta (u_{ss}-u)  {\cal L}_w(\partial z_w) \ , \\
	S_{fl} &=& - \frac{k}{L}\int dt \int dx_4 \int_{u_T}^{+\infty}  du \, \Theta (u_{ss}-u) \Theta (u-u_{ff}) {\cal L}_{fl}(\partial z_{fl}) \ , \\
	S_{sf}& =& - \frac{2k}{L}\int dt \int dx_4\, \delta(x_4-L/2) \int_{u_T}^{+ \infty} du  \int^{z_{fl}(u,L/2)}_{z(\hat C)} dz{\cal L}_d  \ , \\
	S_{h} &=&  -\frac{k}{L} \int dt \int dx_4 \int_{u_T}^{+\infty} du\, \delta(u-u_{J}) \int_{z(\hat C)}^{z(\tilde C)}dz \,{\cal L}_{ch}\ .
	\ea
\end{subequations}
In this case the trailing wall, with Lagrangian ${\cal L}_w(\partial z_w)$, extends from the horizon to the curve $C$ at $u=u_{ss}$, reaching the position $u_w$, with coordinate $z_w$.
There it glues to a ``forward leaning'' part of the profile, with Lagrangian ${\cal L}_{fl}(\partial z_{fl})$,\footnote{${\cal L}_w(\partial z_w)$ and ${\cal L}_{fl}(\partial z_{fl})$ are formally the same, but the boundary conditions are different.} which extends from $C$ to the curve $\tilde C$ at $u=u_{ff}$, with coordinate $z_{fl}$.
At $\tilde C$ the ``forward'' part joins the connected configuration at $u_J$.
In order to obtain the correct brane configuration, we have subtracted the two slides below the trailing wall from the disconnected configuration (as before), and the two ``slides for the forward part'' (``sf'') below the ``forward part'' from the connected configuration.
These slides extend in $u$ from $u_J$ to the ``forward'' profile, and in $z$ from $z(\hat C)$ (where the connected configuration terminates at the curve $\hat C$) to $z_{fl}(u)$. 
Finally, we have subtracted the horizontal part of the connected configuration, with Lagrangian ${\cal L}_{ch}$, extending from $z(\hat C)$ to $z(\tilde C)$.

The domain of the trailing profile $z=z_w(u,x_4)$ is the region $D_C$ spanned by $x_4 \in [-L/2,L/2]$ and $u \in [u_T,C]$, where $C$ is the curve $u=u_{ss}$.
The domain of the forward profile $z=z_{fl}(u,x_4)$ is the region $D_f$ spanned by $x_4 \in [-L/2,L/2]$ and $u \in [u_J,C]$.
  
Taking the variation of the action with respect to $z_w (u), z_f (u)$, we get
\ba
\d S &=& \frac{1}{L} \int dt \int dx_4 \int_{u_T}^{+\infty}  du\,  \Theta (u_{ss}-u) [\p_u \pi_\xi ^u \, \d z_w (u) +  \Theta (u-u_{ff}) \p_u \pi_f ^u \, \d z_{fl} (u)]\nb \\
&-& \frac{1}{L} \int dt \int dx_4 \int_{u_T}^{+\infty} du\, \delta(u-u_{ss}) [-k \mc{L}_c \d z (\hat C) + \pi_\xi ^u \d z_w (C) + \pi_{fl} ^u \d z_{fl} (C)]  \nb \\
&+&\frac{1}{L} \int d t \int dx_4 [\pi_\xi ^u |_{u=u_T} \d z_w (u_T) + \pi_f ^u |_{u=u_J} \d z_{fl} (\tilde C)] \nb \\
&-& \frac{1}{L} \int d t \int_{u_T} ^{+\infty} du \pq{ k \mc{L} _d}_{x_4=L/2} \d z_w (u,L/2) \nb \\
&+& \frac{1}{L} \int d t \int_{u_T} ^{+\infty} du \pq{ - k  \mc{L} _d}_{x_4=-L/2} \d z_w (u,-L/2)   \nb \\
&+& \frac{1}{L} \int d t \int_{u_J} ^{u_w} du k \mc{L} _d \pq{\d z_{fl} (u,L/2) - \d z_{fl} (\hat C)}_{x_4=L/2}  \nb \\
&-& \frac{1}{L} \int d t \int_{u_J} ^{u_w} du k  \mc{L} _d \pq{ \d z_{fl} (u,-L/2) - \d z_{fl} (\hat C) }_{x_4=-L/2} \nb \\
&+& \frac{1}{L} \int dt \int dx_4 \int_{u_T}^{+\infty} du\, \delta(u-u_{J}) k \mc{L}_{ch} [\d z_{fl} (\tilde C) -\d z (\hat C)] 
\ .
\ea

These variations, considering that $\d z_w (C)= \d z_{fl} (C)=\d z (\hat C)$,\footnote{Indeed, the two parts of the configuration, the leaning and the forward walls, join at $C$, so the variation must be the same there. Moreover, $\hat C$ is just the projection at $u_J$ of $C$, so the variation in $z$ must be the same for the two curves.} give the equations
\be\label{eqsal}
\partial_u\pi_{\xi}^u =   2 k {\cal L}_d \,, \qquad \partial_u\pi_{fl}^u =  - 2 k {\cal L}_d \,, 
\ee
and the boundary conditions
\bea \label{bcal}
&& 0 = \pi_{fl}^u (u_J) + k \mc{L}_{ch} (u_J)\,, \nonumber \\ 
&& 0 = \int_{u_T}^{+\infty} du \left[\delta(u-u_w) \left( \pi_{\xi}^u (u) + \Theta(u-u_J) \pi_{fl}^u (u) - k \mc{L}_{ch} (u) \right) + \delta(u-u_J) k \mc{L}_{ch} \right] \nonumber \\
&& \quad + 2k \int_{u_T}^{\infty} du \mc{L} _d \,. 
\eea 
The second boundary condition can be ultimately rewritten as
\be \label{bcal2}
\pi_{\xi}^u (u_w) = -  \pi_{fl}^u(u_w)\,.
\ee 
Integrating equations (\ref{eqsal}) on their domain and using the relations (\ref{bcal}), we obtain
\be
\overline{\pi_{\xi}^u(u_T)} \equiv \frac{1}{L}\int dx_4\, \pi_{\xi}^{u}(u_T, x_4)  =  A\, \Delta p = A (p_t - p_{f})\,.
\ee

Integrating now explicitly equations (\ref{eqsal}), we get
\be
\pi_{\xi}^u = \frac47 \frac{k}{L} R^{3/2} ( u^{7/2} - u_*^{7/2})\ , \qquad \pi_{fl}^u = - \frac47 \frac{k}{L} R^{3/2} ( u^{7/2} - \tilde u^{7/2})\ .
\ee
Using the boundary conditions we get 
\be
u_*^{7/2} = \tilde u^{7/2} = u_T^{7/2} - \frac{7L}{4k R^{3/2}} \, A\Delta p \,.  
\ee

The trailing wall and the forward wall profiles are now given by
\be 
\xi' =  \frac47 \frac{k R^{3/2}}{L}\left( u^{7/2} - u_*^{7/2}\right)\left(\frac{u}{R}\right)^{-3/2}f_T(u)^{-1/2}\sqrt{\frac{1-f_T(u)^{-1}v^2}{k^2 f_T(u) u^8 - \frac{16}{49} \frac{k^2 R^{3}}{L^2}\left( u^{7/2} - u_*^{7/2} \right)^2}}\,,
\ee 
and
\be
\xi_{fl}' =  - \frac47 \frac{k R^{3/2}}{L}\left( u^{7/2} - \tilde u^{7/2}\right)\left(\frac{u}{R}\right)^{-3/2}f_T(u)^{-1/2}\sqrt{\frac{1-f_T(u)^{-1}v^2}{k^2 f_T(u) u^8 - \frac{16}{49} \frac{k^2 R^{3}}{L^2}\left( u^{7/2} - \tilde u^{7/2} \right)^2}}\,.
\ee 
At the radial position $u_c$ where the numerators and the denominators in the square roots of $\xi', \xi_{fl}'$ are zero, 
one has
\be
k f_T^{1/2} u_c^4 = \frac47 \frac{k R^{3/2}}{L} (u_c^{7/2} - \tilde u^{7/2}) \,,
\ee
which gives the relation providing the velocity as a function of $\Delta p$
\be
\Delta p= \frac{2^5}{3^9} \pi^3 \lambda^3 N N_f  (L T) \frac{T^7}{M_{KK}^3} \left[ \frac{v}{(1-v^2)^{4/3}}  +\frac47 \frac{3}{4\pi L T}\left( 1- \frac{1}{(1-v^2)^{7/6}} \right)\right]\,,
\ee
which is exactly the same as (\ref{effesuahere}).
Thus, all the considerations and the values for the velocity obtained before are still true.

\begin{figure}
	\center
	\includegraphics[scale=0.6]{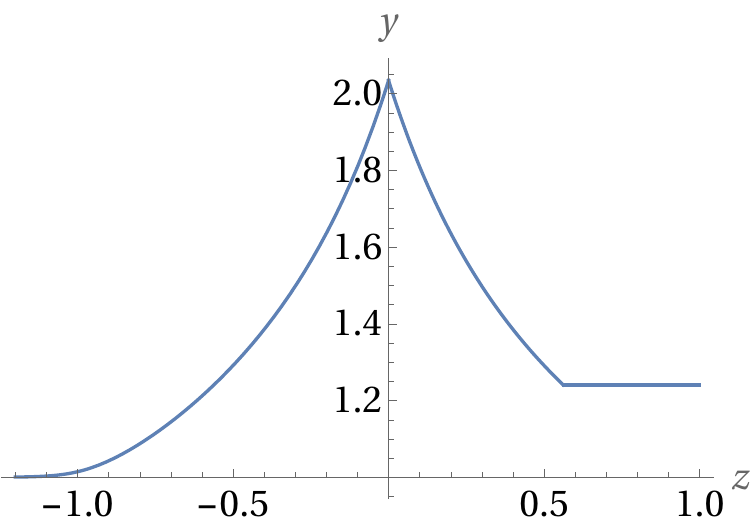}
	\caption{Plot of the alternative brane profile along the holographic radial direction $y$ for $\bar T=1.05$, corresponding to the value of the velocity $v=0.111$.}
	\label{fig:profilealter}
\end{figure}

A sample profile of the trailing wall and the forward leaning wall is reported in figure \ref{fig:profilealter}, corresponding to a fixed-$x_4$ section of the cartoon in figure \ref{fig:fluid} at fixed $t$.

\section{Conclusions}
\label{S:concl}

In this work, we have studied the physics of bubbles in superheated first-order phase transitions in Holographic QCD, namely the Witten-Sakai-Sugimoto model.
We have considered both the confinement-deconfinement transition and the chiral symmetry restoration one.

In the former case, we have employed the effective approach pioneered in \cite{Creminelli:2001th} and used for the supercooled transition in the WSS model in \cite{Bigazzi:2020phm}.
For the chiral symmetry restoration transition, we have employed an improved version of the variational ansatz of \cite{Bigazzi:2020phm} to derive the bubble profile, represented by a specific configuration of the flavor $D8$-branes.  

In both cases, we have derived the bubble profile as a function of $T/T_c$, with $T_c$ the critical temperature of the transition, and from this, the values of the bubble radius and on-shell action, both of which are monotonically decreasing as expected from the phenomenology of first order phase transitions.
We have also computed the bubble nucleation rate for different values of the relevant model parameters, $M_{KK}$ and $g=\lambda N^2$ or $\lambda$ for the deconfined and chiral symmetry restoration cases, respectively. 
It turns out that as these parameters increase, the nucleation rate curves shift to higher temperatures, while the peak value remains unchanged.

When compared to the characteristic scales expected in a specific realization of the first order phase transitions, these results allow to extract the nucleation and percolation temperatures $T_{n,p}$, as well as the inverse transition duration $\beta$.
We have presented the example of transitions in neutron star mergers.
For standard values of the parameters $g$ and $\lambda$, we get for both types of transitions values of $\beta \sim {\cal O}(100)$ and $T_{n,p}/T_c \approx [1.4,1.6]$ (deconfinement), $T_{n,p}/T_c \approx [1.01,1.02]$ (chiral symmetry restoration).
Of course, this application is just an example; the  general formulas can be applied to other scenarios as well that may arise in cosmological phase transitions upon reheating.

For the deconfinement transition, we qualitatively argued that the bubble wall velocity is very small in the holographic limit. On the other hand, for the chiral symmetry restoration transition,  we have followed the more quantitative approach of \cite{Bigazzi:2021ucw} to compute the steady-state bubble wall velocity $v$ in what we called the rectangular approximation.
Although the approximation is rather crude, the results for the friction force and the velocity are quite neat. The friction force comprises two contributions: the drag force, which quantifies the energy loss through the black hole horizon, and another piece given by the difference between the pressure of the true vacuum at temperature $T$ and the same pressure at the boosted temperature. It turns out that this other piece has the opposite sign relative to its counterpart obtained in the supercooled case \cite{Bigazzi:2021ucw}, which was given by the difference between the pressure of the false vacuum at the boosted temperature and the pressure of the false vacuum at temperature $T$. As a result, the bubble wall velocity of the superheated phase transition  is always smaller than its  supercooled counterpart as depicted in figure \ref{fig:velocity}.  This qualitative behavior is consistent with the only previous result from holography \cite{Bea:2024bls}, which used a bottom-up model at finite density. Nevertheless, our maximum  value of the steady-state velocity is rather small yet larger than the one obtained in \cite{Bea:2024bls}.  
Moreover, it turns out that the drag force is proportional to the enthalpy density of the true vacuum rather than that of the false vacuum, as it turned out in the supercooled case.

An obvious future direction to our work is to improve the approximations employed in our study, e.g., by numerically solving the full  equations for the $D8$-brane embedding to compute the bubble nucleation rate and by employing a variational ansatz to determine the steady-state velocity. Of course, a fully-fledged dynamical evolution of the bubble wall would be desirable.

Another obvious future direction is to include a finite chemical potential.
This would, for example, allow for a more faithful application of the model to the physics of neutron stars.
The ultimate goal is to calculate the spectra of gravitational waves associated with these transitions, to have direct benchmark results for observations.
Some of these extensions are currently under study.

\section*{Acknowledgments}
We are deeply indebted to Angel Paredes for his collaboration in the early stages of the project, for his comments, insightful suggestions and help with the numerics. We are grateful to Giulio Barni and David Mateos for inspiring discussions and to Mikel Sanchez-Garitaonandia for interesting observations.
This project has been partially supported by the grant PRIN 20227S3M3B ``Bubble Dynamics in Cosmological Phase Transitions''.

\end{document}